\documentclass[aps,prc,preprint,amsmath,showpacs,superscriptaddress,nofootinbib]{revtex4}
\usepackage{graphicx,epsfig}
\newcommand{\beqy}{\begin{eqnarray}}
\newcommand{\eeqy}{\end{eqnarray}}
\newcommand{\bmlet}{\begin{subequations}}
\newcommand{\emlet}{\end{subequations}}

\begin{document}

\textwidth 16.2 cm
\oddsidemargin -.54 cm
\evensidemargin -.54 cm

\def\gsimeq{\,\,\raise0.14em\hbox{$>$}\kern-0.76em\lower0.28em\hbox
{$\sim$}\,\,}
\def\lsimeq{\,\,\raise0.14em\hbox{$<$}\kern-0.76em\lower0.28em\hbox
{$\sim$}\,\,}

\title{Inner crust of neutron stars with mass-fitted Skyrme functionals}
\author{J.~M.~Pearson}
\affiliation{D\'ept. de Physique, Universit\'e de Montr\'eal, Montr\'eal 
(Qu\'ebec), H3C 3J7 Canada}
\author{N.~Chamel}
\affiliation{Institut d'Astronomie et d'Astrophysique, CP-226, Universit\'e
Libre de Bruxelles, 1050 Brussels, Belgium}
\author{S.~Goriely}
\affiliation{Institut d'Astronomie et d'Astrophysique, CP-226,
Universit\'e Libre de Bruxelles, 1050 Brussels, Belgium}
\author{C. Ducoin}
\affiliation{CFC, Department of Physics, University of Coimbra,
P3004--516, Coimbra, Portugal} 
\date{\today}

\begin{abstract}

The equation of state and composition of the inner crust of neutron stars
at zero temperature are calculated, using the $T$ = 0 version of the 
TETFSI (temperature-dependent extended Thomas-Fermi plus Strutinsky integral) 
method, for each of a family of three functionals based on Skyrme-type forces
BSk19, BSk20 and BSk21, which are characterized by different 
degrees of symmetry-energy stiffness, and also for the SLy4 functional. 
We also solve the Tolman-Oppenheimer-Volkoff equations to calculate the
distribution of mass within the inner crust. Qualitatively similar results
are found for all four functionals, and in particular the number of protons
per Wigner-Seitz cell is in all cases equal to 40 throughout the inner crust.
\end{abstract}


\maketitle

\section{Introduction}

We recall that three distinct regions can be recognized in a neutron star: a 
locally homogeneous core and two concentric shells characterized by different 
inhomogeneous phases~\cite{pr95,lrr}. The outermost of the shells, the 
``outer crust", consists of an electrically neutral lattice of nuclei and 
electrons. At the surface of the star only nuclei that are stable under natural
terrestrial conditions are found (in fact, under the assumption of ``cold 
catalyzed matter", i.e., nuclear and beta equilibrium at temperature $T$ = 0, 
only $^{56}$Fe will be found), but on moving towards the interior the 
increasing density leads to the appearance of nuclei that are more and more neutron rich, 
until at a mean local density $\bar{n}$ of around 2.5 $\times 10^{-4}$ 
nucleons fm$^{-3}$ (4.2 $\times 10^{11}$ g cm$^{-3})$ neutron drip sets in. 
This marks the transition to the ``inner crust", an inhomogeneous assembly of 
neutron-proton clusters and unbound neutrons, neutralized by an essentially
uniform electron gas. By the point where the mean density has risen to about
two thirds of the density $n_0$ of symmetric (homogeneous) nuclear matter (SNM) at 
equilibrium, the 
inhomogeneities have been smoothed out and we enter the core of the star.
The homogeneous medium of which the core is comprised is known as 
``neutron-star matter" (N*M), and is made up primarily of neutrons, with a 
small admixture of protons neutralized by electrons (and muons at densities 
above $\bar{n}\simeq 0.12$ fm$^{-3}$). Closer to the 
center, other particles such as hyperons might appear.

In this paper we continue our calculations of the different regions of neutron
stars with a family of three Skyrme-type functionals,
BSk19, BSk20 and BSk21, that we have constructed specifically to provide a 
unified approach not 
only to the structure of the different regions of neutron stars but also to 
other phenomena associated with the birth and death of neutron stars, e.g., 
supernova-core collapse, the r-process of nucleosynthesis in the neutrino-driven wind,
and nucleosynthesis via the decompression of neutron-star matter~\cite{gcp10}. 
These three functionals are all based on effectives forces with the generalized 
Skyrme form
\beqy \label{1.1}
v_{ij} & = &
t_0(1+x_0 P_\sigma)\delta({\pmb{r}_{ij}})
+\frac{1}{2} t_1(1+x_1 P_\sigma)\frac{1}{\hbar^2}\left[p_{ij}^2\,
\delta({\pmb{r}_{ij}}) +\delta({\pmb{r}_{ij}})\, p_{ij}^2 \right]\nonumber\\
& + & t_2(1+x_2 P_\sigma)\frac{1}{\hbar^2}\pmb{p}_{ij}.\delta(\pmb{r}_{ij})\,
 \pmb{p}_{ij}
+\frac{1}{6}t_3(1+x_3 P_\sigma)\,n(\pmb{r})^\alpha\,\delta(\pmb{r}_{ij})
\nonumber\\
&  + & \frac{1}{2}\,t_4(1+x_4 P_\sigma)\frac{1}{\hbar^2} \left[p_{ij}^2\,
n({\pmb{r}})^\beta\,\delta({\pmb{r}}_{ij}) +
\delta({\pmb{r}}_{ij})\,n({\pmb{r}})^\beta\, p_{ij}^2 \right] \nonumber\\
& + &t_5(1+x_5 P_\sigma)\frac{1}{\hbar^2}{\pmb{p}}_{ij}.
n({\pmb{r}})^\gamma\,\delta({\pmb{r}}_{ij})\, {\pmb{p}}_{ij} 
 + \frac{\rm i}{\hbar^2}W_0(\mbox{\boldmath$\sigma_i+\sigma_j$})\cdot
\pmb{p}_{ij}\times\delta(\pmb{r}_{ij})\,\pmb{p}_{ij}  \quad ,
\eeqy
where $\pmb{r}_{ij} = \pmb{r}_i - \pmb{r}_j$, $\pmb{r} = (\pmb{r}_i +
\pmb{r}_j)/2$, $\pmb{p}_{ij} = - {\rm i}\hbar(\pmb{\nabla}_i-\pmb{\nabla}_j)/2$
(this is the relative momentum), $P_\sigma$ is the two-body spin-exchange
operator, and $n(\pmb{r}) = n_n(\pmb{r}) + n_p(\pmb{r})$ is the total
local density, $n_n(\pmb{r})$ and $n_p(\pmb{r})$ being the neutron and
proton densities, respectively. The $t_4$ and $t_5$ terms here are 
unconventional, being density-dependent generalizations of the $t_1$ and $t_2$ 
terms, respectively. 

\begin{figure}
\centerline{\epsfig{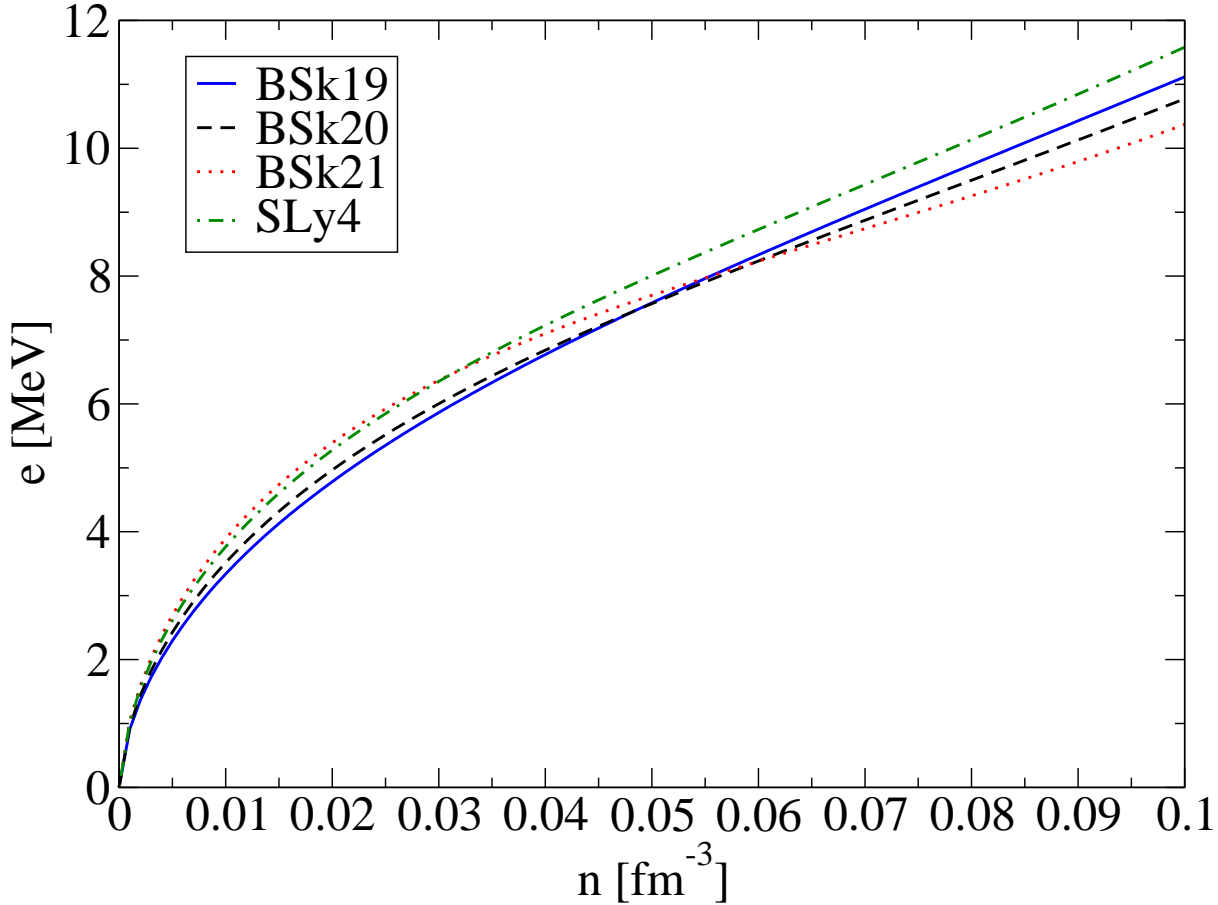}}
\caption{(Color online.) Neutron-matter EOSs (internal energy per nucleon $e$ 
as a function of density $n$) for forces BSk19 -- 21
and SLy4 at subnuclear densities and zero temperature.} 
\label{fig1}
\end{figure}

The parameters of this form of force were determined primarily by fitting 
measured nuclear masses, which were calculated with the Hartree-Fock-Bogoliubov
(HFB) method. For this it was necessary to supplement the Skyrme forces with a 
microscopic contact pairing force, phenomenological Wigner terms and correction
terms for the spurious collective energy. However, in fitting the mass data we 
simultaneously constrained the Skyrme force to fit the zero-temperature 
equation of state (EOS) of homogeneous neutron matter (NeuM), as 
determined by many-body calculations with realistic two- and three-nucleon 
forces; the strength of the pairing force at each point in the nucleus in 
question was likewise calculated analytically so as to reproduce  the $^1S_0$ 
pairing gaps of homogeneous nuclear matter of the appropriate density and 
charge asymmetry~\cite{cha10}. Actually, several realistic calculations of the 
EOS of NeuM 
have been made, and while they all agree very closely at nuclear and subnuclear
densities, at the much higher densities that can be encountered towards the 
center of neutron stars they differ greatly in the stiffness, i.e., the density
dependence, of the symmetry energy that they predict, and there are very few 
data, either observational or experimental, to discriminate between the 
different possibilities. It is in this way that we arrived at the three 
different functionals of this paper: BSk19 corresponds to the softest
EOS of NeuM known to us, BSk21 to the stiffest, while BSk20 has intermediate
symmetry stiffness, as seen in Fig.~1 of Ref.~\cite{gcp10}. 
On the other hand, Fig.~\ref{fig1} of the present paper shows that in        
NeuM the three functionals are very close to each other at the subnuclear     
densities relevant to neutron-star crusts. For a further discussion of        
this point see Ref.~\cite{gcp10}, where it will be seen in particular that    
a value of 30 MeV was imposed on the symmetry coefficient $J$ for all         
three functionals. It will also be seen there that the values of the          
density-symmetry coefficient $L$, which measures the stiffness of the         
symmetry energy {\it at the equilbrium density} $n_0$, are all very           
similar.

Furthermore, we imposed on 
these functionals the supplementary constraints of i) eliminating all 
unphysical instabilities in nuclear matter for all densities up to the maximum 
found in neutron stars (these functionals are also stable at the finite 
temperatures encountered in supernova cores~\cite{cg10}) ii) obtaining a 
qualitatively realistic distribution of the 
potential energy among the four spin-isospin channels in nuclear matter 
iii) ensuring that the isovector effective mass is smaller than the isoscalar 
effective mass, as indicated by both experiment and many-body calculations. 

The introduction of 
the unconventional terms in $t_4$ and $t_5$ allowed us to satisfy all these 
constraints and at the same time fit the 2149 measured masses of nuclei with 
$N$ and $Z \ge$ 8 given in the 2003 AME (Atomic Mass Evaluation)~\cite{audi03} 
with an rms deviation as low as 0.58 MeV for all three models, i.e., for all 
three options for the high-density variation of the symmetry energy. 
For all three of these 
functionals complete mass tables (labeled HFB-19, HFB-20 and 
HFB-21, respectively) were constructed, going from one drip line to the other.
The reliability of the predictions that these models make for experimentally 
inaccessible neutron-rich nuclei is all the greater for the constraints to 
neutron matter imposed on their underlying forces, and it was thus particularly
appropriate to use these mass models in our earlier study of the outer crust of
neutron stars~\cite{pgc11}. 
As for the homogeneous core, the $T$ = 0 EOSs of N*M for our forces BSk19, 
BSk20 and BSk21 have already been published in the original paper presenting 
these forces~\cite{gcp10}. This leaves just the inner crust to be dealt with,
and our main concern in this paper is to calculate for this region the EOS and 
the composition as a function of density with each of our three functionals.

We shall also perform inner-crust calculations with the functional 
SLy4~\cite{cha98}, since like our own functionals it is designed for 
finite-nucleus HFB calculations, and is intended for neutron-star studies, 
being subject to a neutron-matter constraint. However, it has the 
conventional Skyrme form and thus, having fewer parameters, is far less
flexible than our own functionals. Thus SLy4 was fitted to only six nuclear 
masses; moreover, three of these nuclei had $N = Z$ (even), and since no Wigner
term was included in the model the symmetry energy must inevitably be too 
large. (In particular, the symmetry coefficient $J$ for this functional is 32 
MeV, while we have found that the optimal value for the conventional form of  
Skyrme functional when all the mass data are fitted without any neutron-matter 
or other constraint is 28 MeV~\cite{sg05}.) The excessive symmetry energy might 
explain why the rms deviation from the mass data is quite large, 5.1 MeV~\cite{dsn04}; 
note that only even-even nuclei were considered in that calculation.

Given that all four functionals 
were fitted to masses with the HFB method, it might 
seem appropriate to use this method for the inner-crust calculations as well.
Now the latter calculations have been generally performed within the 
framework of the spherical Wigner-Seitz (WS) approximation, as in the pioneer 
HF calculations of Negele and Vautherin~\cite{nv73}, in order to avoid computer
times grossly in excess of those for isolated-nucleus calculations. 
But an inevitable consequence of the WS approximation is to introduce shell 
effects in the spectrum of unbound neutron states, which dominate the 
properties of the inner crust. Such shell effects are to a large extent 
spurious, since in reality the unbound neutron states form a quasi-continuum. 
This difficulty is analyzed in detail in Refs.~\cite{cha07,marg07}, the latter 
reference showing that the error thereby introduced in the energy per nucleon 
cannot easily be reduced below 50 keV, which is incompatible with a reliable 
calculation of the composition of the inner crust; for a very
recent discussion of the problem see Grill {\it et al.}~\cite{gri11}. In the
last few years 3D calculations have been carried out by several
groups~\cite{mag02,gog07,new09}. However, not only does this sort of
calculation require computer times that are quite impractical for extensive
astrophysical calculations but the use of a cubic box with periodic boundary
conditions can still lead to spurious neutron shell effects (see, for
example, Section C.2 in Ref.~\cite{new09}). 

In view of these problems it is not surprising that a more popular approach to 
the calculation of the inner crust has been to use the much simpler 
compressible liquid-drop model (CLDM); a typical such calculation is that of 
Ref.~\cite{dh01}. Within each Wigner-Seitz cell this method makes a clear 
separation of nuclear matter into two distinct homogeneous phases, the 
densities of which are free parameters of the model. The bulk properties of
the two phases are calculated microscopically using the adopted functional,
as are the surface properties (preferably including curvature corrections) of 
the interface between them. A more realistic treatment of spatial 
inhomogeneities is to employ semi-classical methods such as the Thomas-Fermi 
(TF) approximation, as for instance in Ref.~\cite{oya07}. 
However, both the CLDM and TF methods are otherwise purely macroscopic, and in 
particular have no quantum shell corrections at all.

The so-called TETFSI method (temperature-dependent extended Thomas-Fermi plus 
Strutinsky integral) of Onsi {\it et al.}~\cite{ons08}, which we adopt here, is 
a computationally very fast approximation to the full finite-temperature HF 
method. This method was originally developed for calculating the EOS of the 
dense matter found in supernova cores~\cite{ons97}. But in this work we will 
be using just the zero-temperature limit. The TETFSI method, like the TF 
method, allows for a continuous variation of the density of nuclear matter 
within each WS cell, without any artificial separation into two distinct 
phases. However it is expected to provide a much better description of nuclear 
clusters than the TF method because the semi-classical expressions for the 
kinetic-energy and spin current densities include density-gradient terms up to 
the fourth order. Most importantly, proton shell corrections are added 
perturbatively, but we avoid the difficulty of spuriously large values for the
neutron shell corrections noted above by not calculating them at all; in any 
case they are known to be much smaller than the proton shell 
corrections~\cite{oy94,ch06,cha07}. 

Our method is described in detail in Ref.~\cite{ons08}, but we summarize it 
here in Section II. The results for the zero-temperature composition and EOS 
of the inner crust are presented in Section III, along with an examination of 
the extent to which continuity holds at the interface with the outer crust. 
This section also discusses the transition between the inner crust and the 
liquid core. In Section IV we examine the solutions to the 
Tolman-Oppenheimer-Volkoff
(TOV) equations~\cite{tol39,ov39} in order to determine the  distribution of 
mass within the inner crust. Our conclusions are summarized in Section V.   

\section{TETFSI model of inner crust}
\label{model}

To summarize the main features of the TETFSI method~\cite{ons08}, we note
first that it models the inhomogeneous medium by spherical WS 
cells, with the spherically symmetric neutron and proton density distributions 
being parametrized according to 
\beqy\label{2.1}
n_q(r) = n_{Bq} + n_{\Lambda q}f_q(r)  \quad ,
\eeqy
in which, with $q = n$ or $p$, $n_{Bq}$ is a constant background term, while 
\beqy\label{2.2}
f_q(r) = \frac{1}{1 + \exp \left\{\Big(\frac{C_q - R}
{r - R}\Big)^2 - 1\right\} \exp \Big(\frac{r-C_q}{a_q}\Big) }\quad ,
\eeqy
In this ``damped" form of the usual simple Fermi profile all density 
derivatives vanish at the surface of the cell, thereby ensuring a smooth 
matching of the nucleonic distributions between adjacent cells, and satisfying
certain necessary conditions discussed below. It is particularly to be noted 
that with this parametrization of the density there is no arbitrary separation 
into liquid and gaseous phases within the WS cell. However, if this were what 
is energetically 
favored in reality, it would automatically be taken into account
through the small values of the diffusenesses $a_q$ that would emerge. 

In order to determine the composition and the EOS of the inner crust one should
in principle minimize at constant pressure the Gibbs free energy $g$ per 
nucleon with respect to all the parameters of the WS cell. This is the 
procedure that we adopted in Ref.~\cite{pgc11} for the outer crust, but for the
inner crust the computation would be extremely heavy. Instead, we
minimize rather the total Helmholtz free energy $f$ per nucleon at constant
mean density $\bar{n}$ with respect to the same parameters, showing in
Appendix~\ref{minimiz} that the error thereby introduced is quite negligible.
Since the present work is limited to $T$ = 0 it is the internal energy per
nucleon $e$ that is minimized ($f = e - Ts$, where $s$ is the entropy per
nucleon).

To enumerate the minimizing parameters of the WS cell, we note
first that the cell radius $R$ will be determined, for the given $\bar{n}$, 
by the total number $A$ of nucleons in the cell. Then with the number of 
protons $Z$ and the number of neutrons $N$ in the cell specified ($Z + N = A$),
only three of the four remaining cell parameters appearing in Eqs.~(\ref{2.1}) 
and (\ref{2.2}) for each charge-type of nucleon will be independent. Thus, 
including $Z$ and $N$, there will be eight parameters with respect to which 
the  energy $e$ must be minimized. Identifying the different contributions 
to $e$, we write
\beqy\label{2.3}
e= e_{\rm nuc} + e_e + e_c - Y_e\,Q_{n,\beta} \quad ,
\eeqy
and now discuss briefly each term.  

The nuclear term is 
\beqy\label{2.4}
e_{\rm nuc} = \frac{4\pi}{A}\int_{\rm cell} r^2
{\mathcal E}_{\rm Sky}^{\rm ETF}(r)dr + e_{\rm sh}^p   \quad  ,
\eeqy
where ${\mathcal E}_{\rm Sky}^{\rm ETF}(r)$ is the ETF approximation to the 
energy density ${\mathcal E}_{\rm Sky}(r)$ given by Eq.~(A3) of 
Ref.~\cite{cgp09} for the generalized Skyrme force (\ref{1.1}) (the formalism 
of Ref.~\cite{ons08} is limited to conventional Skyrme forces). All terms in
${\mathcal E}_{\rm Sky}$ are functions of the number densities $n_q(r)$, the 
kinetic-energy densities $\tau_q(r)$ and the spin-current densities 
$\pmb{J}_q(r)$. The ETF method approximates these last two densities as 
functions of the number densities $n_q(r)$ and their first four derivatives. 
However, as far as the total ETF
energy is concerned, it is shown in App. A of Brack {\it et al.}~\cite{bgh85} that the
third- and fourth-order derivatives of the density can be eliminated by partial 
integration over the region of interaction, provided certain boundary conditions are 
satisfied on the bounding surface of this region. In the finite-nucleus case of 
Ref.~\cite{bgh85} the bounding surface can be taken to lie at infinity, in which case 
the necessary boundary conditions are trivially easy to satisfy. In the present
case the bounding surface is the surface of the WS cell, and, given the fact
that in general the density does not vanish on this surface, the necessary 
conditions are that the first three derivatives of the density must vanish 
there. These 
conditions are satisfied automatically for the distribution~(\ref{2.2}), and it is in
this ``integrated" form that we have implemented the ETF method. Note that for 
all four functionals, i.e., BSk19--21 and SLy4, we omit the spin-current terms in 
$J^2$, since this is the way these functionals were fitted (see Ref.~\cite{cg10} for a
discussion of the implications of these terms for spin and isospin stability).   

With the ETF approximations for $\tau_q(r)$ and $\pmb{J}_q(r)$ being
semi-classical all shell effects in ${\mathcal E}_{\rm Sky}(r)$ are lost. The 
second term on the right-hand side of Eq.~(\ref{2.4}) represents our attempt to
restore the proton shell corrections perturbatively using the SI (Strutinsky
integral) method, as described in Ref.~\cite{ons08}. As explained in Section I,
we do not calculate neutron shell corrections in the inner crust; for a fuller 
discussion of this point see Section I of Ref.~\cite{ons08}, where we conclude
that because of the problems with neutrons, the ETFSI method is better adapted
to a WS approach than is the HFB (or HF-BCS) method. On the other hand, we do
not include pairing at the present stage of our calculations. This should have
very little impact on the EOS, but it might have implications for the 
composition. 

The term $e_e$ on the right-hand side of Eq.~(\ref{2.3}) denotes the kinetic 
energy per
nucleon of the electrons. In dense, cold, neutron star crust, electron-charge 
screening effects are negligible and the electron density $n_e = \bar{n}_p$ is 
essentially uniform~\cite{wat03,mar05}. The energy $e_e$ can thus be calculated 
straightforwardly by expressions given in Section 24 of Cox and 
Giuli~\cite{cg04}, as in Ref.~\cite{ons08}. 

The third term on the right-hand side of Eq.~(\ref{2.3}) denotes the total 
Coulomb energy per nucleon. It is calculated according to Eq.~(3.4) of
Ref.~\cite{ons08}, except that there are the following changes to the exchange
part. a) The proton exchange energy is set equal to zero for the three BSk 
functionals; this is a device that
we have successfully adopted in all our recent models, beginning with 
BSk15~\cite{gp08}, and it can be interpreted as compensating for neglected 
effects such as Coulomb correlations, charge-symmetry breaking of the nuclear 
forces, and vacuum polarization. b) The electron exchange energy, which has the 
nonrelativistic form in Eq.~(3.4) of Ref.~\cite{ons08}, is multiplied by a 
factor of -1/2, as appropriate for extremely relativistic 
particles~\cite{sal61}.
 
The last term on the right-hand side of Eq.~(\ref{2.3}), in which $Q_{n,\beta}$
is the beta-decay energy of the neutron (0.782 MeV) and $Y_e = Z/A$, takes 
account of the neutron-proton mass difference (we drop a constant term 
$M_n c^2$).

Minimization of $e$ with respect to the eight available parameters is performed
by means of the CERN routine MINUIT. Actually, we found it necessary to exclude
the shell correction term $e_{\rm sh}^p$ from this minimization, and then to add it
later to what is really just the optimal ETF part of the energy. Otherwise, 
the minimization routine will tend to seek large negative values of the shell 
corrections, in violation of the essentially perturbative character of the SI 
method. In practice, we performed the
minimization for different fixed values of $Z$, thereby reducing the number of
free variational parameters to seven. However, even with this reduced number 
of parameters MINUIT occasionally failed to find a correctly converged minimum.
This problem could often, but not always, be avoided by adjusting the initial 
values for the parameters. When this procedure failed solutions could always be
found, provided we are not too close to the interface with the core, by
a slight shift in the value of $\bar{n}$; for this reason our grid of values of
$\bar{n}$ is irregular. However, above a certain value of $\bar{n}$ we were
unable to find any solutions at all when MINUIT minimizes with respect to 
seven variables. We attribute the failure of our code to find well defined 
minima before true homogeneity has been reached to the energy minimum being 
very flat, with the result that MINUIT is unable to pick out one configuration 
among a very wide range of possibilities. We found, however, that we could
still find well defined minima in this region by reducing the number of free 
variables in MINUIT to three, $n_{\Lambda n}, n_{\Lambda p}$ and $N$, 
and minimizing for a large number of fixed
values of the other five parameters; clearly, for a given level of accuracy 
this procedure will require much more computation time than when MINUIT 
minimizes on seven variables. 

It should be noted that at all densities the number of neutrons $N$ in       
the WS cell is taken as one of the minimizing variables in MINUIT and         
hence is treated as a continuous variable, rather than being discretized      
to integral values. Even though the total number of neutrons in the           
crustal layer is, of course, integral, the notion of a fractional number of    
neutrons per WS cell corresponds, in fact, to the physical reality, since     
the neutrons are delocalized.

Normally we would expect positive values of the constants $n_{\Lambda n}$ and 
$n_{\Lambda p}$ to emerge from the minimization, the cluster then representing 
a ``droplet". However, there have been indications~\cite{pr95,lrr} that 
towards the interface with the core the clusters may take several other 
forms. Most of these ``pasta" configurations, such as slabs, tubes and rods,
cannot be handled by our code, which is restricted to spherical shapes, but
another of these possibilities, spherical bubbles, could in principle
emerge from the minimization with our code, since they correspond simply to  
negative values of $n_{\Lambda q}$. We return to this possibility in
Section~\ref{core}.

The pressure $P$ corresponding to any given value of $\bar{n}$ is calculated
by evaluating a simple analytic expression, as described in
Appendix~\ref{pressure}. This is more reliable and computationally much faster
than the numerical differentiation of $e$ used in Ref.~\cite{ons08}.

\section{Composition and equation of state of inner crust}
\label{inner crust}

\subsection{Generalities}

Following the methods described in the previous section, for each of our three 
functionals and SLy4 we minimized the internal energy per nucleon $e$
at temperature $T$ = 0 for more than a hundred different densities $\bar{n}$ 
between the drip point and 0.1 fm$^{-3}$. At this upper limit our density
distributions have become effectively homogeneous, as will be discussed in
more detail in Section~\ref{core}.

For all values of $\bar{n}$ up to 0.06 fm$^{-3}$ the optimal value of the 
number of protons $Z$ per Wigner-Seitz cell was found to be 40, for all four 
functionals. However, at higher densities, as homogeneity is approached, the 
minimized energy becomes increasingly insensitive to $Z$. The preference 
for $Z$ = 40 in the case of these four functionals is somewhat fortuitous, 
given that for some of our older functionals different values were found. For 
example, with the functional BSk14 used in Ref.~\cite{ons08} it was found that 
$Z$ could take any of the values, 20, 40 and 50, according to the density. 
It is remarkable that these familiar finite-nucleus
magic proton numbers should persist in the highly neutron-rich environment
beyond the drip line, especially in view of the presence of electrons, which
will have the effect of significantly reducing Coulomb effects.
 
\begin{figure}
\centerline{\epsfig{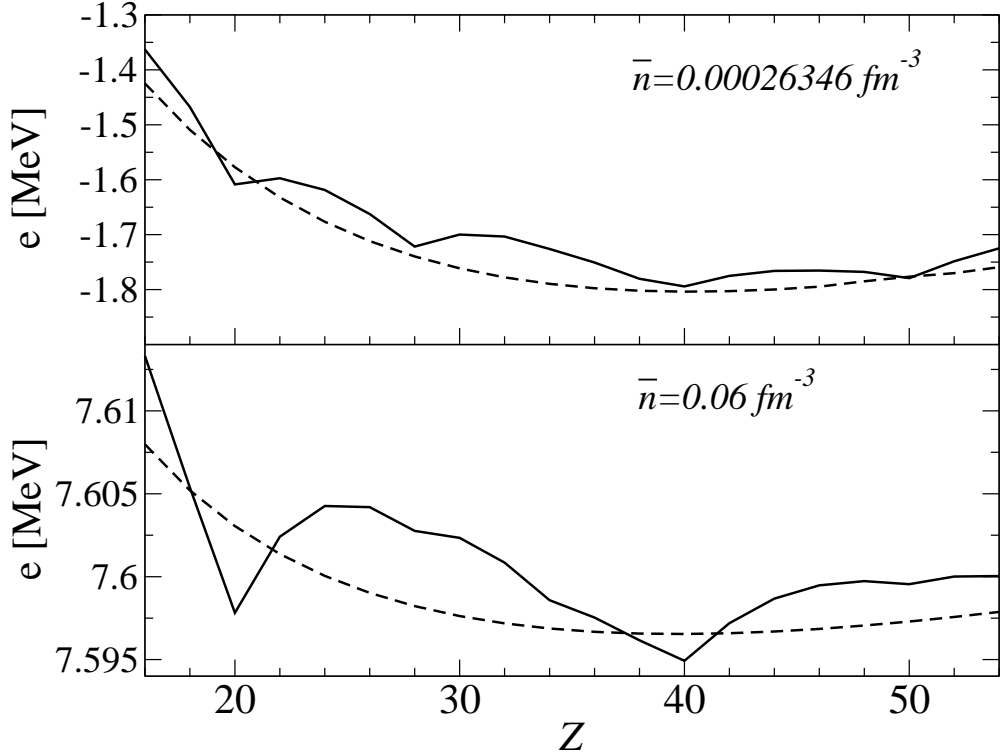}}
\caption{Variation of ETFSI energy $e$ per nucleon as a function of $Z$
for functional BSk19 with $N$ optimized for each value of $Z$; dotted curve
represents ETF approximation. Upper panel: 
$\bar{n} = 2.63 \times 10^{-4}$ nucleons fm$^{-3}$ (drip density); lower 
panel: $\bar{n} = 0.06$ nucleons fm$^{-3}$.}
\label{fig2}
\end{figure}

For the specific case of functional BSk19, reference to Fig.~\ref{fig2}
shows both the role of shell effects and the overall trends imposed by the
ETF part of the calculation. For both of the extreme densities shown here
the ETF minimum lies close to $Z$ = 40, and the shell effects simply
reinforce this preference. However, the energy difference per nucleon
$\Delta\,e$ between $Z$ = 40 and $Z$ =50 is very small: about 10 keV
at the drip density and 5 keV at $\bar{n} = 0.06$ fm$^{-3}$ (note the different
energy scales of the two panels). It is easy to see how with even an only very
slightly different functional 
a quite different $T$ = 0 composition could be found,
as a result of changes in either the shell effects or the macroscopic ETF part
of the energy (or both).

Since the functionals BSk19, BSk20 and BSk21 give better and wider data fits 
than all our earlier functionals, and have a better theoretical base as well, 
we believe
our prediction of $Z$ = 40 at all densities in the inner crust to be more
credible than our earlier predictions, but the need for caution is evident.
For example, taking pairing into account might well shift the favored value of 
$Z$ away from 40. In any case, in a real neutron star a fairly wide range of 
values of $Z$ can be expected at any point in the inner crust
because of the finite temperature.

\begin{figure}
\centerline{\epsfig{figure=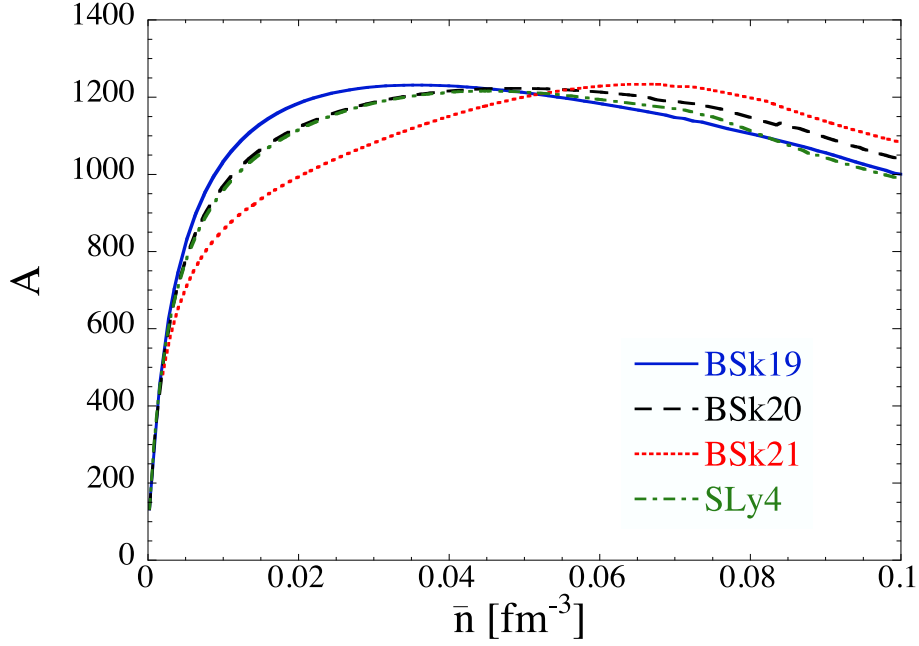,width=13cm,angle=0}}
\caption{Optimal value of nucleon number $A$ as a function of density 
$\bar{n}$ at zero temperature in inner crust; proton number $Z$ everywhere 
takes optimal value of 40 for all four forces.} 
\label{fig3}
\end{figure}

\begin{figure}
\centerline{\epsfig{figure=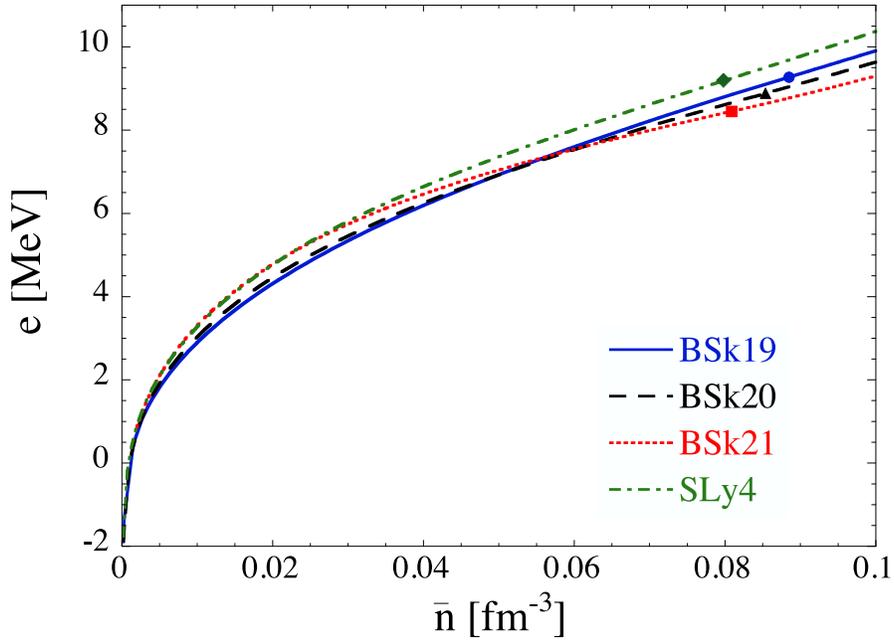,width=13cm,angle=-0}}
\caption{Internal energy $e$ per nucleon at zero temperature as a function of density 
$\bar{n}$ in inner crust. The solid symbols represent the transition densities
$n_{\rm trans}^{N*M}$ (see Section~\ref{core}).}
\label{fig4}
\end{figure}

\begin{figure}
\centerline{\epsfig{figure=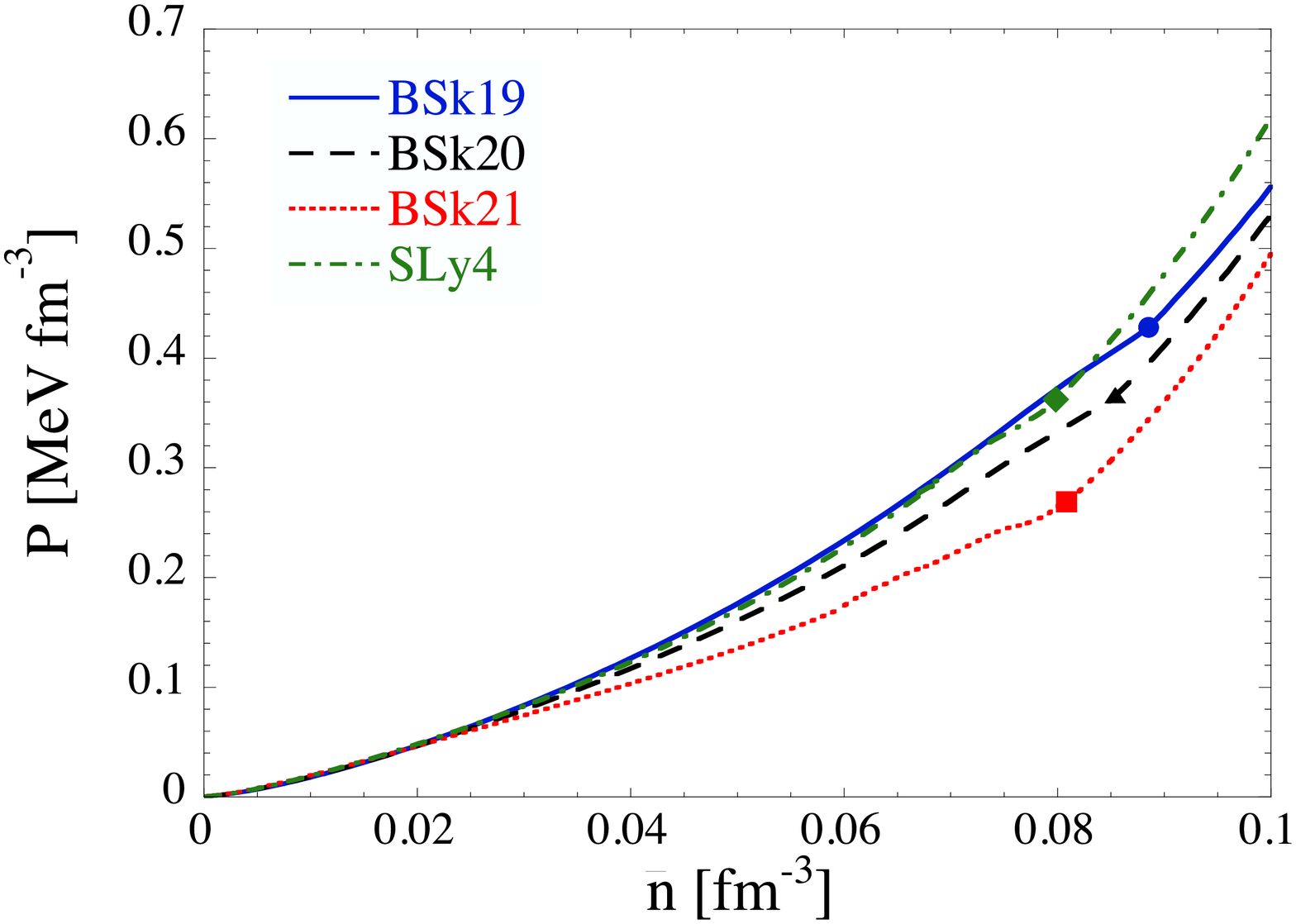,width=13cm,angle=-0}}
\caption{Pressure $P$ per nucleon at zero temperature as a function of density
$\bar{n}$ in inner crust. The solid symbols represent the transition densities
$n_{\rm trans}^{N*M}$ (see Section~\ref{core}).}
\label{fig5}
\end{figure}

The optimal values of $A$ are plotted as a function of the density in
Fig.~\ref{fig3}; similarly, Figs.~\ref{fig4} and \ref{fig5} show the variation
of $e$ and the pressure $P$, respectively; these two figures show
the densities $\bar{n}_{\rm trans}^{N*M}$ of transition between the inner crust and
the core, as calculated in Section~\ref{core}.

No essential differences will be perceived between any of these four 
functionals, as far as the inner crust is concerned, although BSk21 is seen in 
Fig.~\ref{fig5} to have a somewhat softer EOS (in contrast to a much stiffer
EOS at high density). These features can be related to the behavior of the 
respective functionals 
in homogeneous NeuM at inner-crust densities (see Fig.~\ref{fig1}). Since SLy4 
gives a much worse mass fit than do any of the BSk functionals, one might have 
expected that it would represent less well the presence of inhomogeneities and 
of protons, and thus give significantly different results in the inner crust, 
but this turns out not to be the case. Furthermore, the higher value of the 
symmetry coefficient $J$ in the case of SLy4 (32 MeV, as opposed to 30 MeV in 
the case of all the BSk functionals) does not seem to have much impact. 

\subsection{Continuity with outer crust.}

Our inner-crust code, as used here, is in principle applicable to the outer 
crust, with the background densities $n_{Bq}$ vanishing automatically on
minimizing the energy per nucleon, and it is thus
meaningful to compare this code with the code we used for the
outer-crust calculation of Ref.~\cite{pgc11}. In Table~\ref{tab1} we make
this comparison at the drip-point density $\bar{n}_{drip}$ (as determined by 
the code for the outer crust) with the results for the outer-crust code shown
in parentheses. 

We see that the inner-crust code (TETFSI) underbinds with respect to the 
outer-crust code (HFB) by around 5 \%. This disagreement 
can be accounted for by the several approximations
made in our TETFSI method, relative to the HFB method adopted in our 
outer-crust calculations~\cite{pgc11}, as follows. i) The kinetic energy and
spin currents are
calculated with the semiclassical (T)ETF method. ii) Proton shell 
corrections are put in perturbatively, and neutron shell corrections (shown to 
be much smaller than proton shell corrections as soon as neutron drip sets
in~\cite{ch06,cha07}, but obviously 
not zero, in the outer crust) are neglected completely. iii) Rather than
allowing arbitrary density variations when minimizing the total energy, the
density is parametrized according to Eqs.~(\ref{2.1}) and (\ref{2.2}). 
iv) Pairing is neglected completely. We have checked that the assumption of sphericity 
in the inner-crust code has a negligible impact in this region of the nuclear chart.

\begin{table}
\centering
\caption{Comparison of inner-crust and outer-crust codes at drip point;
results for latter code in parentheses. $e$ is the internal energy per
nucleon, and $P$ the pressure.}
\label{tab1}
\vspace{.5cm}
\begin{tabular}{|c|c|c|c|c|c|}
\hline
Force &$\bar{n}_{\rm drip}$ (fm$^{-3}$) &$Z$ & $N$ & $e$ (MeV)& $P$ (MeV fm$^{-3}$)\\
\hline
BSk19 &2.63464$\times10^{-4}$& 40 (38)& 96 (88)& -1.79426 (-1.87464)&5.072$\times10^{-4}$ (4.938$\times10^{-4}$)\\
BSk20 &2.62873$\times10^{-4}$& 40 (38)& 95 (88)&-1.79451 (-1.87305)&5.064$\times10^{-4}$ (4.923$\times10^{-4}$)\\
BSK21 &2.57541$\times10^{-4}$& 40 (38)& 94 (86)&-1.81718 (-1.90057)&4.984$\times10^{-4}$ (4.894$\times10^{-4}$)\\
SLy4 &2.45897$\times10^{-4}$&  40 (38)& 93 (82)&-1.78801 (-1.95898)&4.744$\times10^{-4}$  (4.807$\times10^{-4}$)\\
\hline
\end{tabular}
\end{table}

It will also be seen from  Table~\ref{tab1} that there is a slight 
disagreement in the values of $Z$ and $N$ at the drip point. One might 
speculate that the favoring of $Z$ = 40 over 38 is the result of an 
exaggerated shell effect, but if we drop the proton shell corrections 
altogether then we find slightly higher values of $Z$, typically 41.
However, we have already remarked how the inclusion of 
pairing might well shift the unique value of $Z$ (at $T$ = 0) away from 40, 
and we see from  Fig.~\ref{fig2} that {\it a priori} it would be difficult to
rule out any value of $Z$ between 36 and 50 at the drip density.
The disagreement in the neutron number $N$ is somewhat larger, presumably
because of our neglect of neutron shell effects, but it is $Z$ that is the
more astrophysically relevant nucleonic number.

\subsection{Transition to homogeneous core}
\label{core}

The densities $n_{\rm trans}^{N*M}$ shown in Figs.~\ref{fig4} and \ref{fig5}, 
and tabulated in 
Table~\ref{tab2}, are the densities below which homogeneous beta-equilibrated 
N*M is calculated, for the functional in question, to be unstable to breakup 
into inhomogeneities. Our values for $n_{\rm trans}^{N*M}$ were calculated by the 
method described in Ref.~{\cite{duc07}, in which one defines a free-energy 
curvature matrix by
\beqy
\label{3.1}
C_{\rm NMe,dyn} &=&
\left(
\begin{array}{ccc}
\frac{\partial\mu _{n}}{\partial n _{n}} & \frac{\partial\mu _{n}}{\partial n _{p}} & 0\\
\frac{\partial\mu _{p}}{\partial n _{n}} & \frac{\partial\mu _{p}}{\partial n _{p}} & 0\\
0 & 0 & \frac{\partial\mu _{e}}{\partial n _{e}}\\
\end{array}
\right)
+
k^2
\left(
\begin{array}{ccc}
2C_{nn}^{\nabla} & 2C_{np}^{\nabla} & 0\\
2C_{pn}^{\nabla} & 2C_{pp}^{\nabla} & 0\\
0 & 0 & 0\\
\end{array}
\right)
+
\frac{4\pi^2e^2}{k^2}
\left(
\begin{array}{ccc}
0 & 0 & 0\\
0 & 1 & -1\\
0 & -1 & 1\\
\end{array}
\right) \quad ,
\eeqy
where the $\mu_i (\equiv \frac{\partial f}{\partial n_i}$) are the neutron, 
proton and electron chemical potentials; note that $n_e = \bar{n}_p$.
The coefficients $C_{ij}^{\nabla}$ account for the density-gradient terms
in the nuclear density functional, which come into play in the presence of
inhomogeneities. The third term on the right-hand side of this equation 
gives the Coulomb contribution. Stability of N*M against breakup (actually,
against density fluctuations of infinitesimally small amplitude) will be 
assured as long as the curvature matrix $C_{\rm NMe,dyn}$ has no negative 
eigenvalues for all real values of $k$, the wavenumber of density 
fluctuations. Thus in practice one calculates the lowest eigenvalue of 
$C_{\rm NMe,dyn}$ along the line of $\beta$-equilibrium of N*M in the 
$n_n - n_p$ plane and determines the density $n^{N*M}_{\rm trans}$ at which it 
changes sign.
Along with $n_{\rm trans}^{N*M}$ , Table~\ref{tab2} also shows 
the value of the proton fraction $Y_e$ and the pressure at the transition point.

It is instructive to see how our density distributions, as given by
Eq.~(\ref{2.1}), approach homogeneity as the density increases. In 
Fig.~\ref{fig6} we follow the approach to homogeneity by showing the
neutron and proton density profiles within the WS cell for different values
of the mean density $\bar{n}$. As far as can be seen from this figure, the 
transition to homogeneous matter is very smooth, with no evidence of any
discontinuity. However, it is not clear in this figure at what precise density
homogeneity can be said to set in, but Figs.~\ref{fig7} and \ref{fig8} complete 
the picture in this respect. The former shows the variation of the ``cluster 
strength'' parameters $n_{\Lambda n}$ and $n_{\Lambda p}$ as a function of 
density: Eq.~(\ref{2.1}) shows that homogeneity corresponds to these 
parameters being equal to zero. Now in Fig.~\ref{fig7} we see that, for all 
functionals, these parameters vanish when the density is very close to 
$n_{\rm trans}^{N*M}$, calculated as described above by a completely different
method. A similar conclusion can be drawn from Fig.~\ref{fig8}, where we plot
a more global measure of the departure from homogeneity, the 
``inhomogeneity factor''
\beqy\label{3.2}
\Lambda = \frac{1}{V_{cell}}\int d^3{\bf r}\left(\frac{n({\bf r})}{\bar{n}}
-1\right)^2 \quad ,
\eeqy
where $V_{\rm cell}$ is the volume of the WS cell and the integration goes over 
the cell. We plot this as a function of density in Fig.~\ref{fig8}, where
the transition to homogeneity at a density very close to the density 
$n_{\rm trans}^{N*M}$ is again apparent. 

We stress also that in both 
Figs.~\ref{fig7} and \ref{fig8} the fall to zero of the appropriate measure of
inhomogeneity is smooth, with no indication of any discontinuity. We cannot
exclude the possibility that the transition is first order, albeit very weak,
but all our results are consistent with the transition being of second order or
higher. 

Figs.~\ref{fig6} and \ref{fig7} make it clear that for none of the four 
functionals considered here have we found a spherical bubble configuration 
anywhere in the inner crust. That is, energy minimisation always leads to a 
droplet configuration until homogeneity is reached. This result confirms, as 
far as force SLy4 is concerned, the CLDM calculations of Douchin and 
Haensel~\cite{dh00}. However, we cannot exclude the possibility of very shallow
bubbles in a very narrow density range, although such configurations would be 
of limited astrophysical interest. Note, moreover, that since our calculations 
are limited to spherical configurations we can say nothing about non-spherical 
bubbles, such as the very shallow ones found for SLy4 in Ref.~\cite{gog07}.

\begin{figure}
\centerline{\epsfig{figure=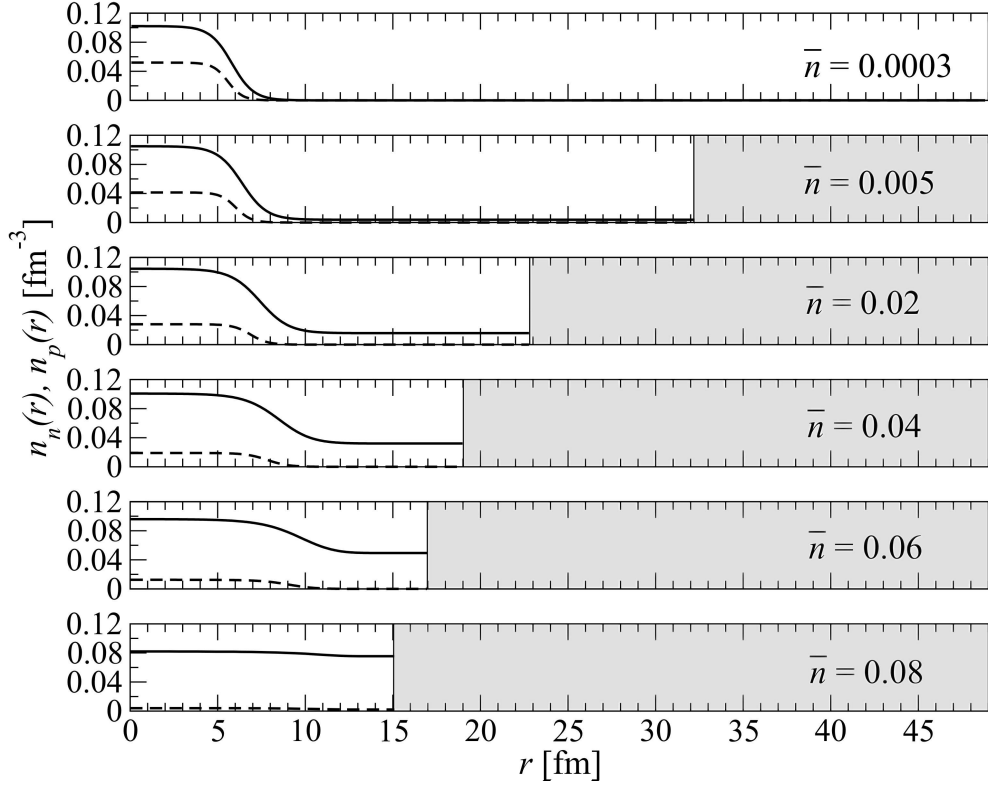,width=13cm,angle=-0}}
\caption{Profiles of neutron (solid curves) and proton (dashed curves) density
distributions in the Wigner-Seitz cell for functional BSk21 and different
values of the mean density $\bar{n}$. Shading denotes the region beyond the
cell radius.}
\label{fig6}
\end{figure}

\begin{figure}{t}
\centerline{\epsfig{figure=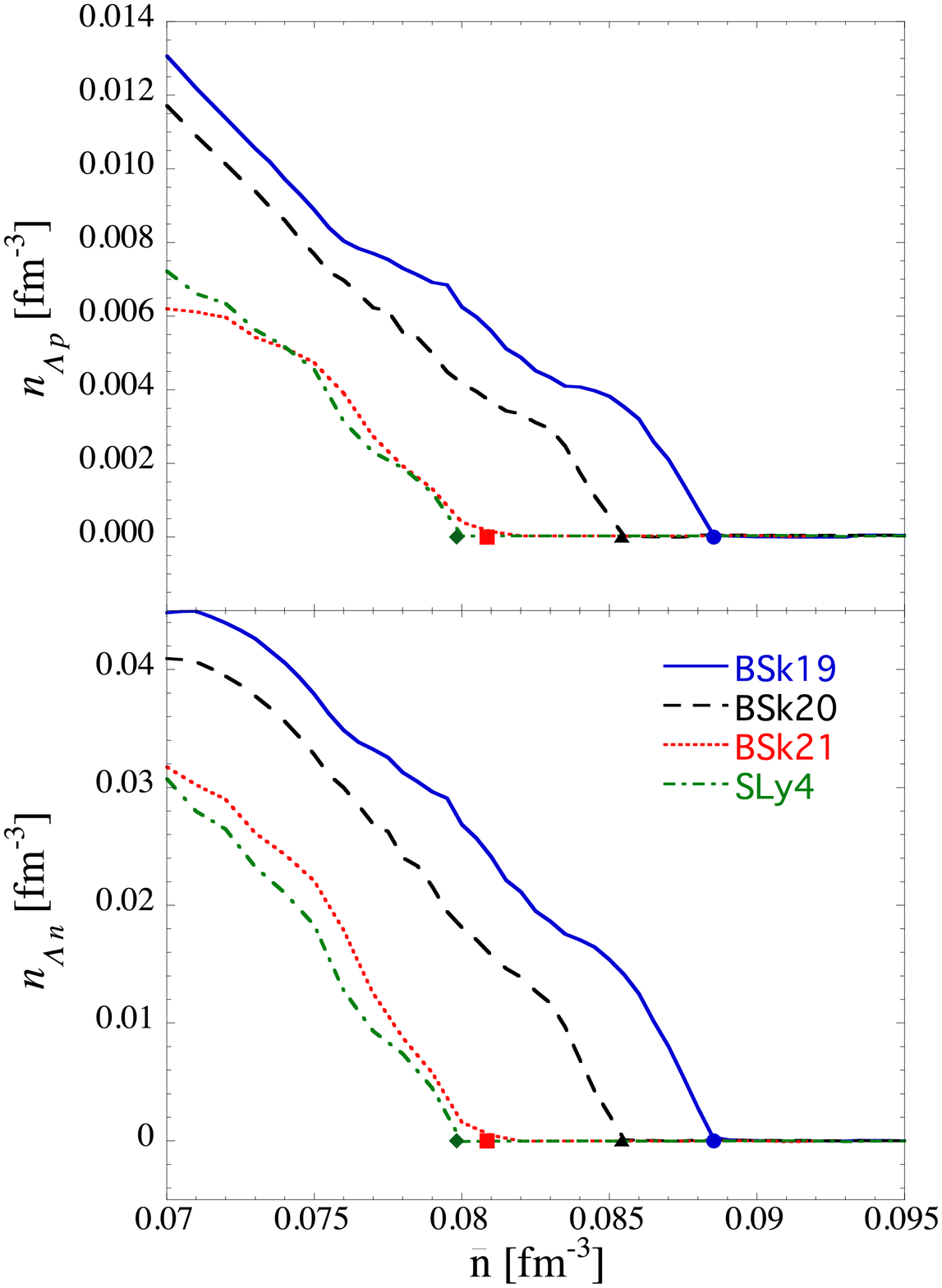,width=13cm,angle=-0}}
\caption{Variation of the ``cluster strength" parameters $n_{\Lambda n}$
and $n_{\Lambda p}$ as a function of density (see Eq.~(\ref{2.1})). 
The solid symbols represent the transition densities
$n_{\rm trans}^{N*M}$ (see Section~\ref{core})}.
\label{fig7}
\end{figure}

\begin{figure}{t}
\centerline{\epsfig{figure=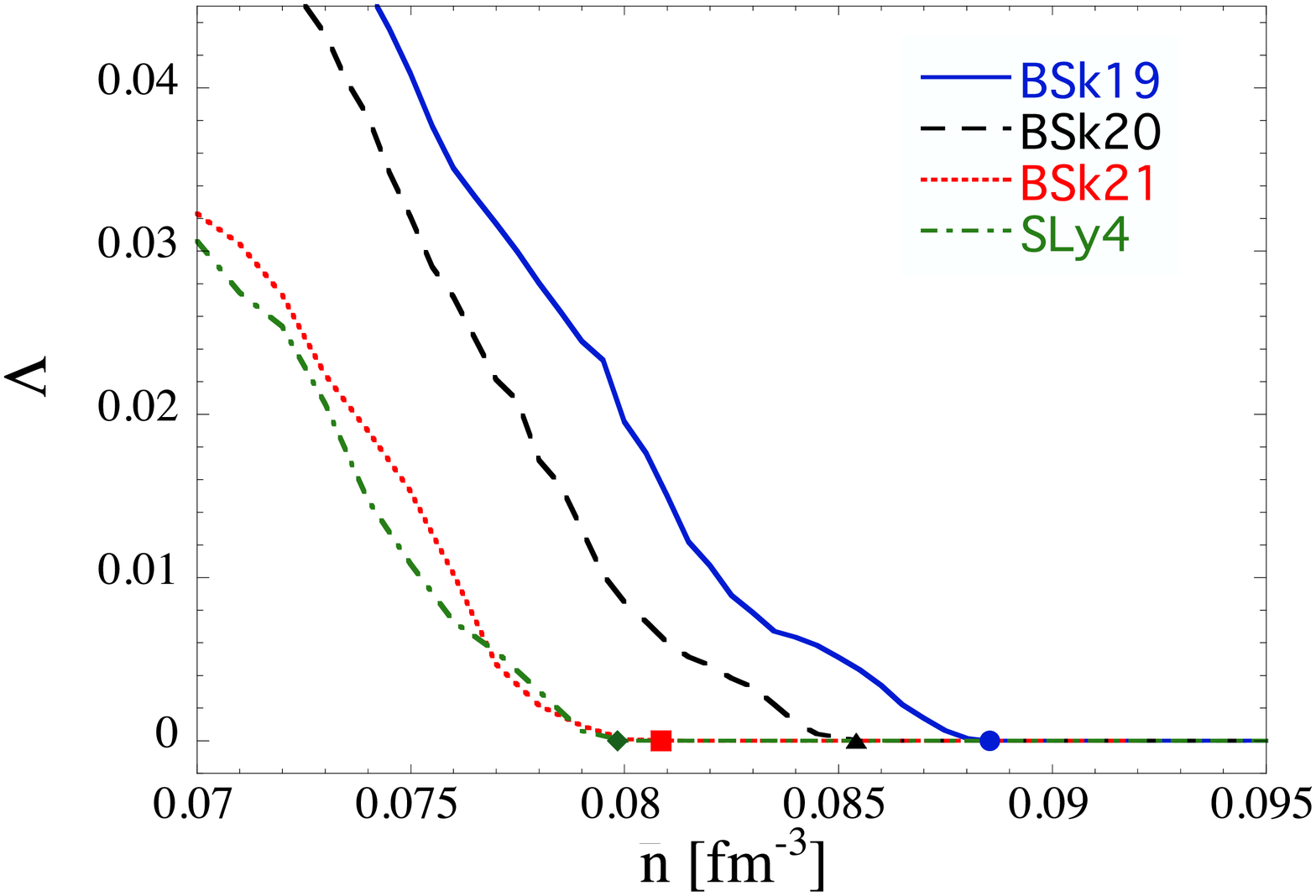,width=13cm,angle=-0}}
\caption{Variation of the inhomogeneity factor $\Lambda$, given by
Eq. (\ref{3.2}), as a function of density. 
The solid symbols represent the transition densities
$n_{\rm trans}^{N*M}$ (see Section~\ref{core})}.
\label{fig8}
\end{figure}

\begin{table}
\centering
\caption{Parameters relating to the crust-core transition.}
\label{tab2}
\vspace{.5cm}
\begin{tabular}{|c|ccc|}
\hline
\hline
Force &$n_{\rm trans}^{N*M}$ (fm$^{-3}$)&$Y_e$ &$P_{\rm trans}$ (MeV fm$^{-3}$)\\
\hline
BSk19 &0.0885&0.0376&0.428\\
BSk20 &0.0854&0.0356&0.365\\
BSk21 &0.0809&0.0335&0.268\\
SLy4  &0.0798&0.0358&0.361\\
\hline
\end{tabular}
\end{table}

\section{Distribution of mass}

With the EOS determined (for a given functional), the distribution of 
mass within a neutron star (assumed to be non-rotating) is given by the 
solution to the TOV equations~\cite{tol39,ov39},  
\beqy
\label{4.1}
\frac{{\rm d}P(r)}{{\rm d}r} = -\frac{G\rho(r)\mathcal{M}(r)}{r^2}
\biggl[1+\frac{P(r)}{c^2\rho(r)}\biggr]
\biggl[1+\frac{4\pi P(r)r^3}{c^2\mathcal{M}(r)}\biggr]\biggl[1-\frac{2G\mathcal{M}(r)}{c^2 r}\biggr]^{-1}
\eeqy
and
\beqy
\label{4.2}
\mathcal{M}(r) = 4\pi\int_0^r\rho(r')r'^2{\rm d}r'\quad .
\eeqy
Here $\rho(r)$ is the mass-energy density at the radial coordinate $r$, given
by 
\beqy\label{4.3}
\rho(r) = \bar{n}(r)\left(M + \frac{e}{c^2}\right)   \quad  ,
\eeqy
where $M$ is the nucleon mass and $e$ is the internal energy per nucleon,
as plotted in Fig.~\ref{fig4}. The pressure $P(r)$ appearing in Eq.~(\ref{4.1})
has to be expressed in terms of $\rho(r)$ through the EOS.

Proceeding as in Section IIIC of Ref.~\cite{pgc11}, the TOV equations 
(\ref{4.1}) and (\ref{4.2}) are solved for the functions $\rho(r)$ and 
$\mathcal{M}(r)$ by integrating inwards from the surface (if we had followed
the usual procedure of integrating outwards from the center our crust results
would have been contaminated by the uncertainties in the EOS of the core). Then
the total baryonic mass of the shell of inner radius $r$ and outer radius $R$, 
the radius of the star, is
\beqy\label{4.4}
\Delta M_B(r) = 4\pi\,M
\int_{r}^{R} r^{\prime 2} {\Phi(r^\prime)}^{1/2}n(r^\prime)dr^\prime  \quad ,
\eeqy
where we have introduced the metric function
\beqy\label{4.5}
\Phi(r) = \left(1 - \frac{2G\mathcal{M}(r)}{c^2r}\right)^{-1} \quad .
\eeqy
Note that $\Delta M_B(r)$, as defined by Eq.~(\ref{4.4}), contains the baryonic
mass of the entire outer crust, as calculated in Ref.~\cite{pgc11} for the 
three BSk forces, and from Refs.~\cite{hempel,rhs06} for SLy4.

We plot $\Delta M_B(\bar{n}(r))$ as a function of the density $\bar{n}$ in 
Fig.~\ref{fig9} for a neutron star of mass 1.5 $M_\odot$ and radius 13 km; the 
fraction of this mass that consists of protons can be
read off from Fig.~\ref{fig3}, given that everywhere we have $Z$ = 40.

For many purposes it might be more convenient to express $\Delta M_B$ as
a function of the proper depth, given by (see Section 5.6 of Ref.~\cite{st83})
\beqy
\label{4.6}
z(r) = \int_r^R {\rm d}r^\prime \left(1 - \frac{2G\mathcal{M}(r^\prime)}
{c^2 r^\prime}\right)^{-1/2}\quad ,
\eeqy
which is the only measurable depth in the gravitationally distorted metric.
We plot in Fig.~\ref{fig10} $\bar{n}$ as a function of $z$, again for a neutron
star of mass 1.5 $M_\odot$ and radius 13 km, whence $\Delta M_B(r)$ can be
read off from Fig.~\ref{fig9} as a function of $z$. 

In Fig.~\ref{fig11} we show how the total gravitational mass of the crust 
(inner plus outer) varies as a function of the total star mass for a given 
radius of 9 km. Fig.~\ref{fig12} shows the same function for stars of radius
14 km. 

\begin{figure}
\centerline{\epsfig{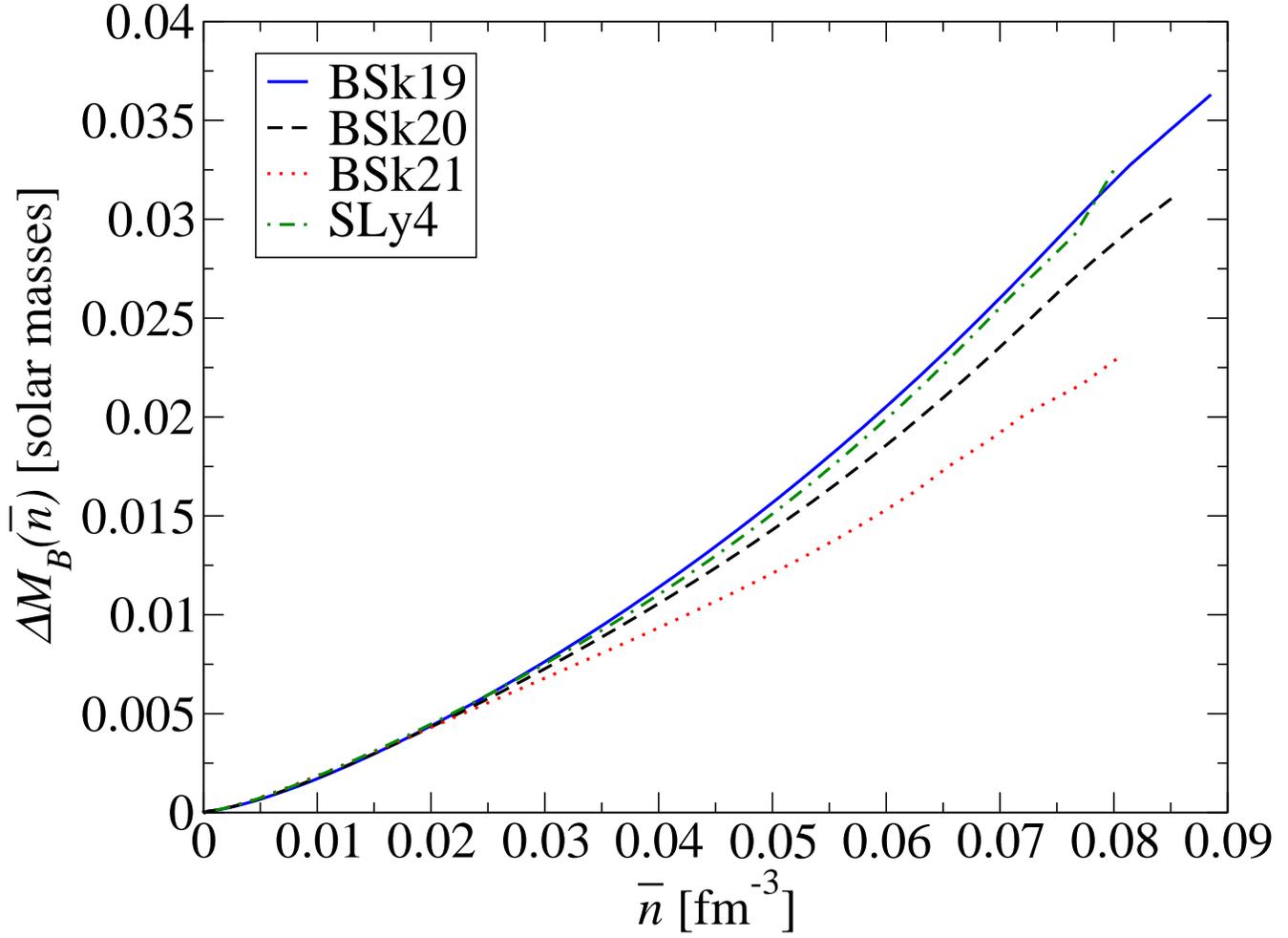}}
\caption{Variation of baryonic mass of crust (inner plus outer) with
density $\bar{n}$ for neutron star of mass 1.5$M_\odot$ and radius 13 km.}  
\label{fig9}
\end{figure}

\begin{figure}
\centerline{\epsfig{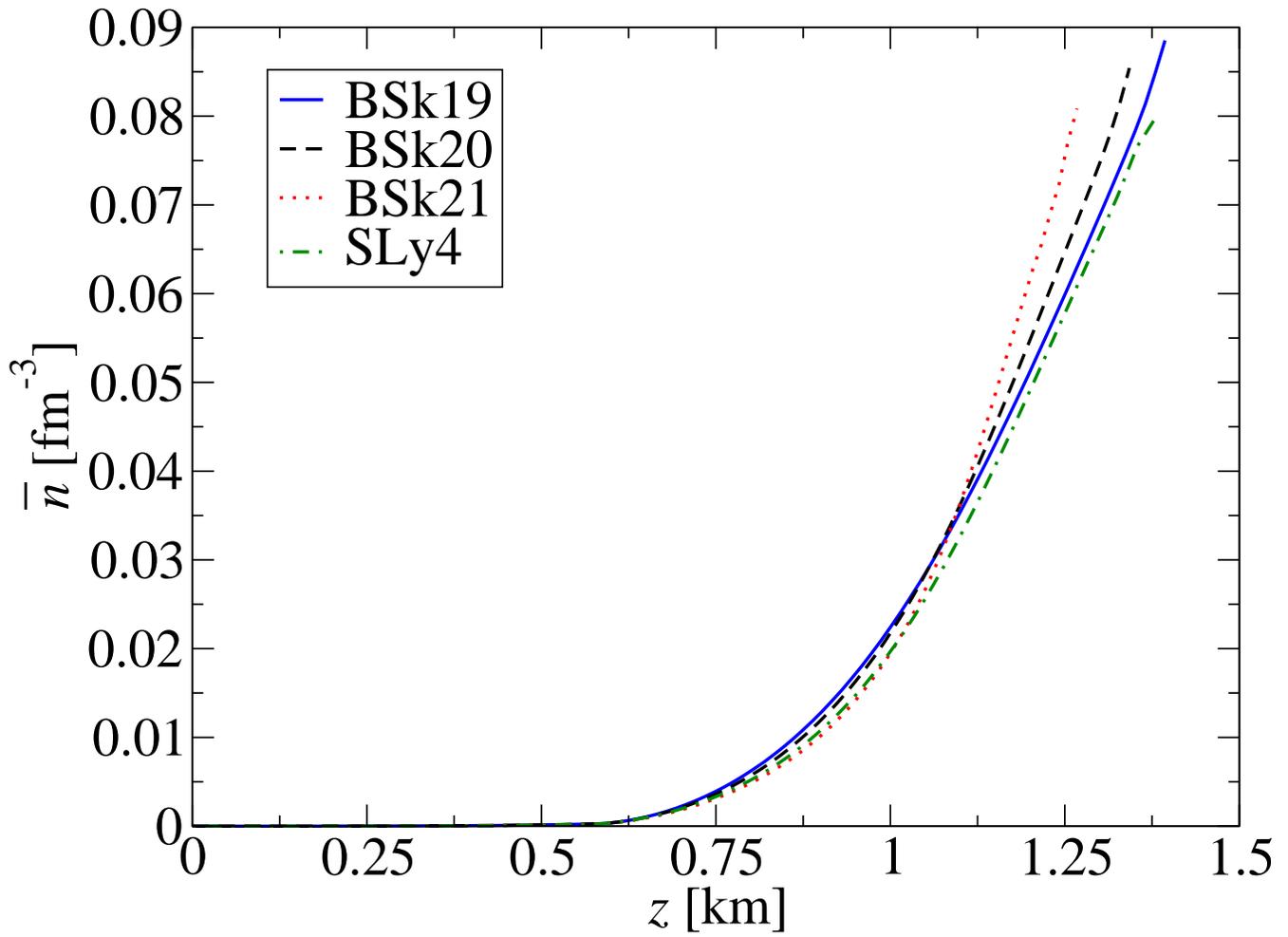}}
\caption{Variation of density $\bar{n}$ with proper depth $z$
for neutron star of mass 1.5$M_\odot$ and radius 13 km.} 
\label{fig10}
\end{figure}

\begin{figure}
\centerline{\epsfig{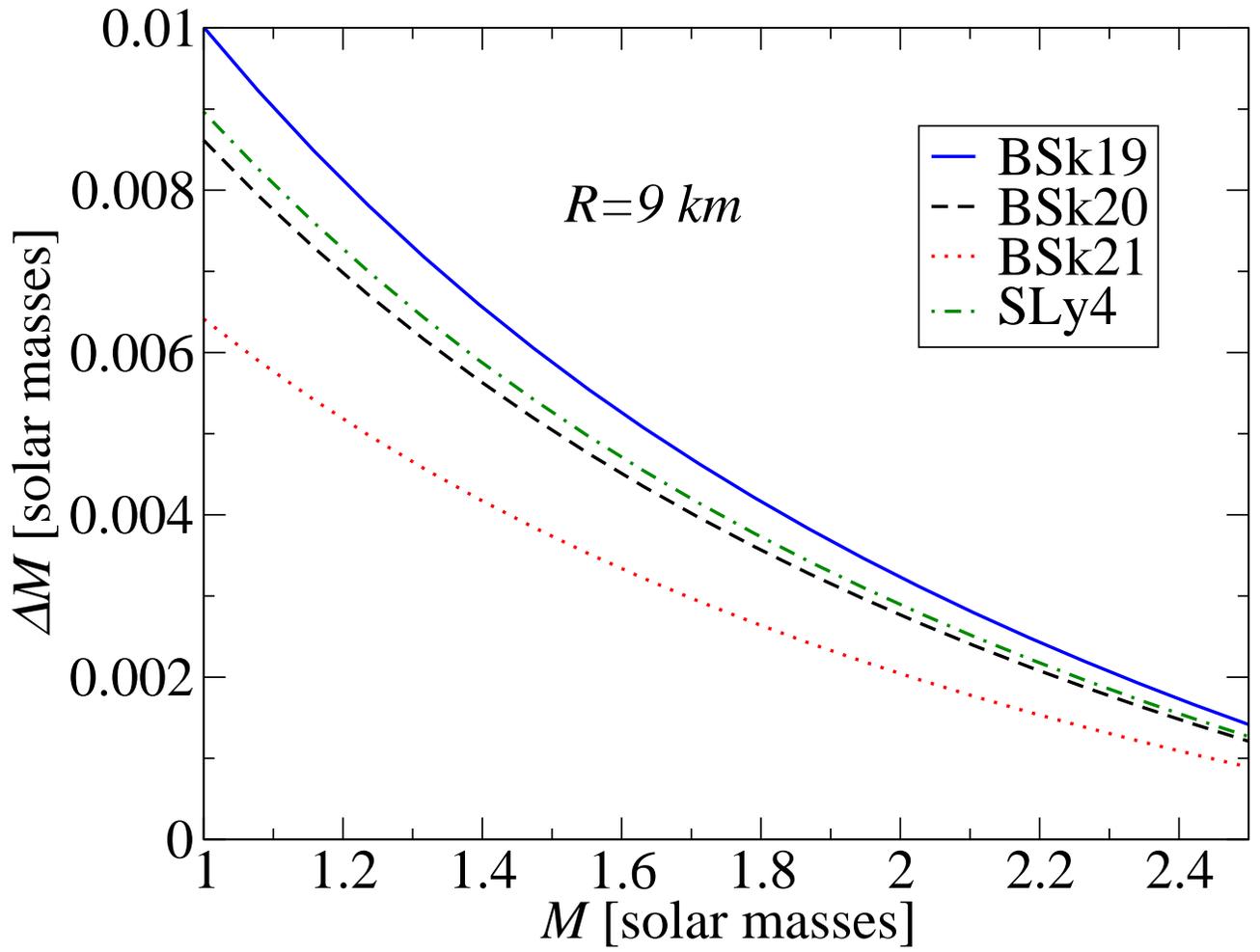}}
\caption{Variation of gravitational mass of crust (inner plus outer)
with total mass of star, radius 9 km.}  
\label{fig11}
\end{figure}

\begin{figure}
\centerline{\epsfig{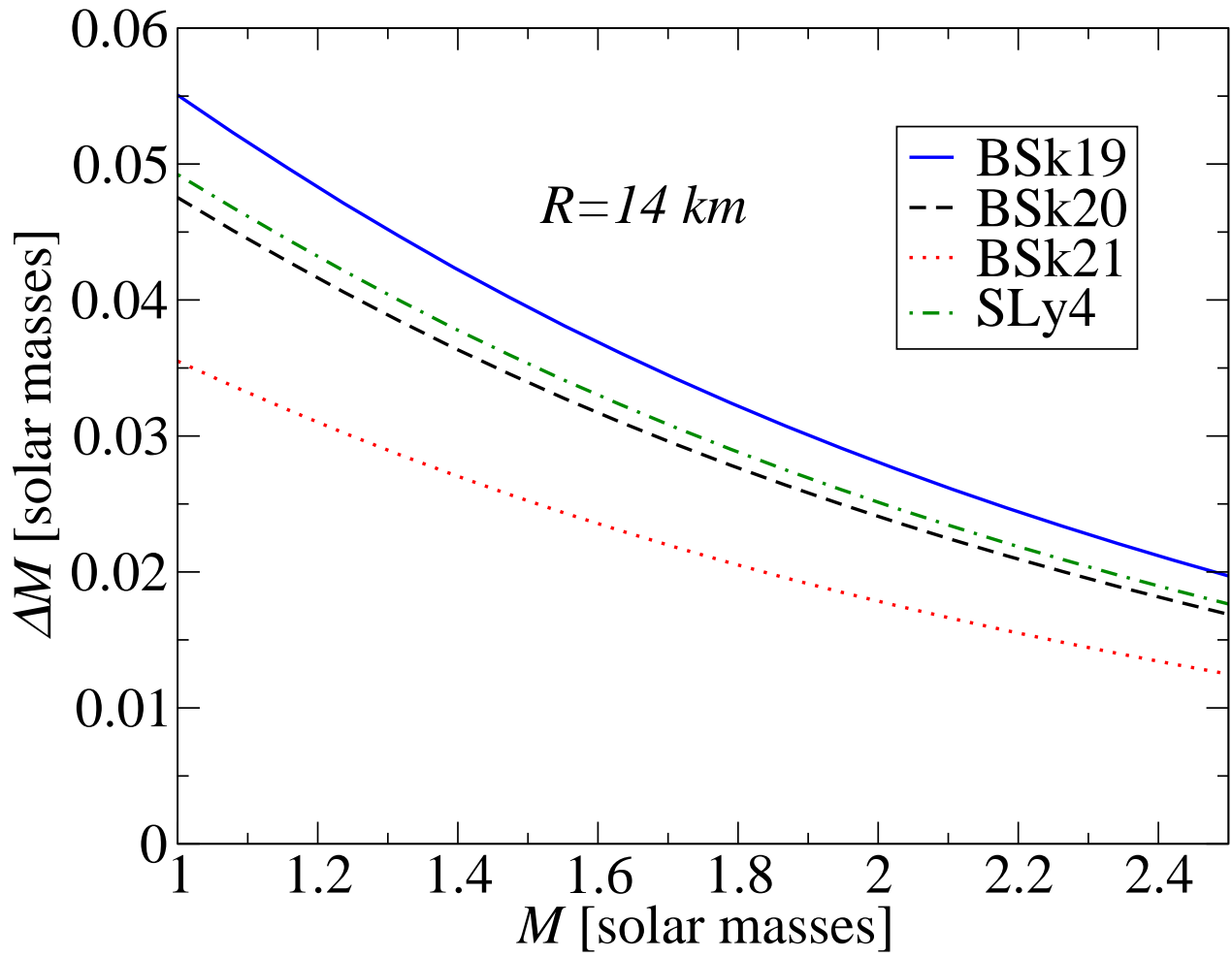}}
\caption{As in Fig.~\ref{fig11}, for stars of radius 14 km.}  
\label{fig12}
\end{figure}

\section{Conclusions}

We have calculated the composition and EOS of the inner crust of neutron stars
for the three generalized Skyrme-type functionals, BSk19, BSk20 and BSk21, and 
for the conventional Skyrme functional SLy4, using in all cases the TETFSI 
method at temperature $T$ = 0. We have also solved the TOV equations to 
calculate the distribution of mass within the crust.

Qualitatively similar results are obtained for all four 
forces. In particular, in all cases we find $Z$ = 40 for the optimal number of
protons per Wigner-Seitz cell throughout the inner crust. However, other 
values of $Z$ lie very close in energy, and if we took pairing into account the
optimal value of $Z$ might very well be shifted away from 40. Moreover, it is 
clear that at realistic values of the 
temperature an appreciable range of values of $Z$ will be found. This
underlines the importance of extending the present calculations to finite
temperatures and to include pairing.

The fact that there are no substantial differences in the 
inner-crust properties for force SLy4 and for the three BSk forces despite
their having been fitted to different values of the symmetry coefficient $J$ 
means that this parameter is not of any great relevance in this respect. 

We have studied in some detail the transition between the inner crust and     
the homogeneous core, considering two different measures of the                
inhomogeneity of our density distributions. We find for each of the four       
functionals that homogeneity is established in our calculated                  
distributions at a density very close to the value predicted for the onset     
in homogeneous N*M of instability against density fluctuations of              
infinitesimally small amplitude. 

No evidence for bubbles was found in the course of this study of the transition
region, despite a thorough search. This conclusion does not preclude the 
existence of non-spherical pasta configurations, a possibility that lies beyond
the scope of the present paper. Even though such phases would have a 
negligeable impact on the EOS, they might affect transport properties.

The calculations on the inner crust presented here show that our forces
BSk19, BSk20 and BSk21 make possible a unified and realistic treatment of all 
regions of neutron stars, as in Ref.~\cite{cha11}.
 
\appendix
\section{Minimization of Gibbs or Helmholtz functions?}
\label{minimiz}

For simple systems, which in the present context means systems with a single 
$(N, Z)$ configuration, minimizing the Gibbs free energy per nucleon $g$ at
constant pressure $P$ is completely equivalent to minimizing the Helmholtz free 
energy per nucleon $f$ at constant density $\bar{n}$, since in that case  the 
thermodynamic identity
\beqy\label{B.1}
\left(\frac{\partial g}{\partial X}\right)_{P,T} =
\left(\frac{\partial f}{\partial X}\right)_{\bar{n},T}  \quad ,
\eeqy
holds, $X$ denoting any thermodynamical variable. But when two different phases
or components, i.e., two different $(N, Z)$ configurations in the present 
context, coexist in equilibrium this identity breaks
down, and it is the Gibbs prescription that leads to a correct description of
the phase transition: there is a discontinuity in the range of densities over
which single-phase solutions can be found, but the pressure remains constant
over this discontinuity, which corresponds to the equilibrium coexistence of
the two phases. If on the other hand one minimizes $f$ at constant density
$\bar{n}$, discontinuities in the pressure will be found in the vicinity of 
transitions from one $(N, Z)$ configuration to another. An example of this
is seen in Fig.~\ref{fig13}, where we show the transition from $Z$ = 40 to $Z$ 
= 20 for functional BSk14~\cite{ons08}, with $N$ being optimized in each case. 
\begin{figure}
\centerline{\epsfig{figure=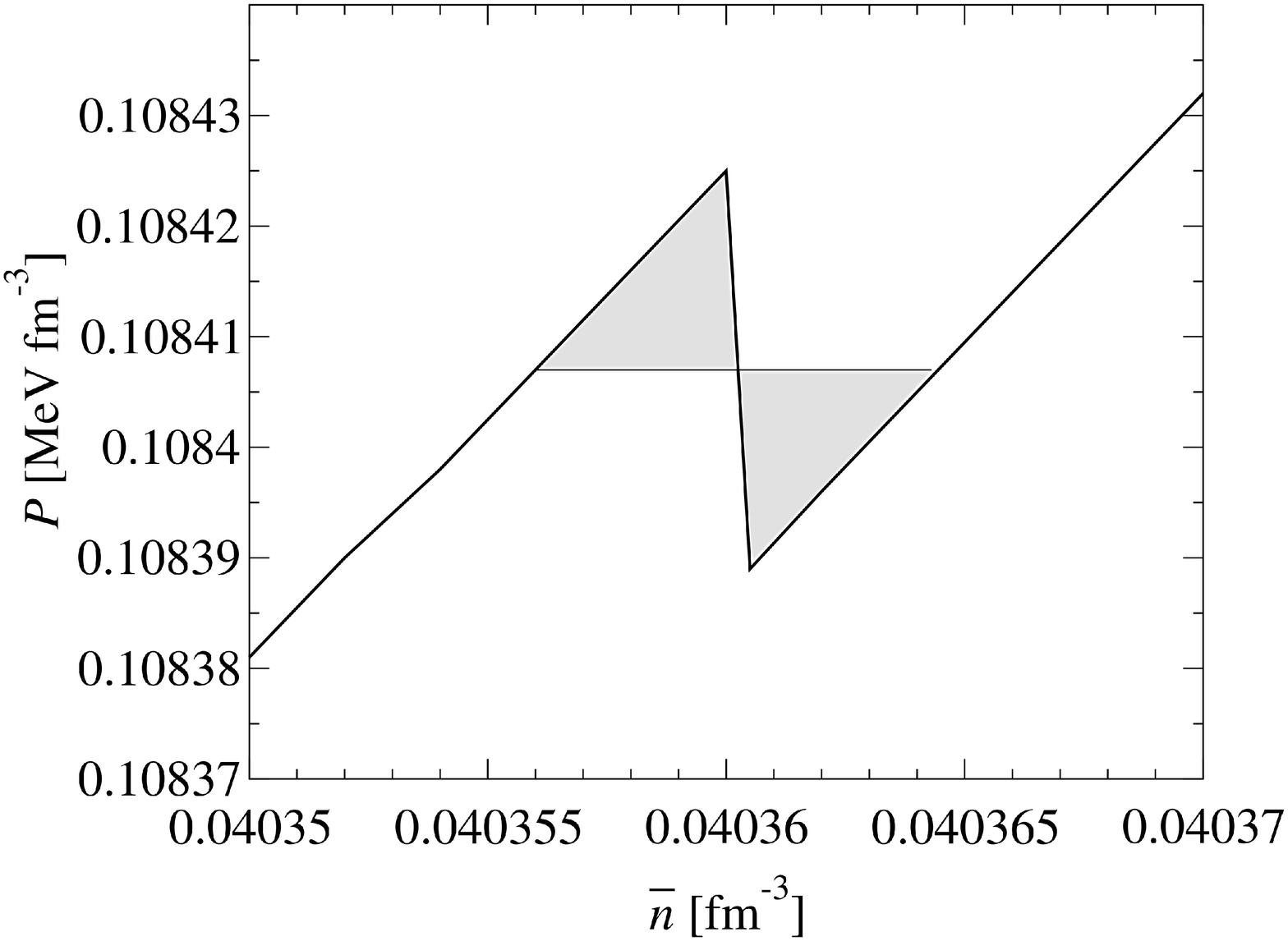,height=13.0cm,angle=-0}}
\caption{EOS for functional BSk14 in the vicinity of the $Z$ = 40 to $Z$ = 20 
transition.}
\label{fig13}
\end{figure}

Such discontinuities in the pressure are unphysical, and arise in our 
calculations only because our model does not allow the coexistence of two 
different $(N, Z)$ configurations that can occur in reality. But even then,
when minimizing $f$ at constant density $\bar{n}$, the correct equilibrium
pressure can be found by making a Maxwell construction, as indicated in 
Fig.~\ref{fig13}. However, on the pressure scale of Fig.~\ref{fig5} these 
discontinuities will be imperceptible, and the Maxwell construction is quite 
unnecessary: the attendant error will be far smaller than the differences
between the EOSs of the different functionals seen in Fig.~\ref{fig5}.

In any case, the question of transitions between different values of $Z$ does 
not arise with functionals BSk19 -- 21, since for all these forces $Z$ retains
the constant value of 40 throughout the inner crust. As for changes in $N$,
we recall that this varies continuously in our calculations, whence it follows 
that minimizing $f$ at constant density $\bar{n}$ leads to absolutely no error 
at all in this respect.
 
\section{Pressure formula}
\label{pressure}

The pressure $P$ at any given point in the neutron-star crust, as given by the 
EOS and as used in the TOV equations, is defined thermodynamically by 
considering a region of volume $V$ that contains the point in question, and is 
macroscopically sized but small enough for all intensive thermodynamic 
variables to be sensibly constant over it. If $F$ denotes the total Helmholtz 
free energy contained in this region then 
\beqy\label{B1}
P = -\left(\frac{\partial F}{\partial V}\right)_{T,N_e,N_q} \quad ,
\eeqy
where $T$ is the temperature (here $T=0$), $N_e$ its number of electrons and 
$N_q$ its number of nucleons of type $q=n,p$ for neutrons and protons, 
respectively. Treating the crust as a perfect crystal, this expression remains 
exact if the region of volume $V$ is taken as the appropriate Wigner-Seitz 
cell, because of the translational symmetry. In the approximation used here of 
spherical WS cells we then have  
\beqy\label{B2}
P =-\frac{1}{4\pi R^2}\left(\frac{\partial F}{\partial R}\right)_{T,N_e,N_q} 
\quad ,
\eeqy
where $R$ is the cell radius. We assume that the Helmholtz free energy in the 
cell can be written as 
\beqy\label{B3}
F=4\pi \int_0^R {\rm d} r\, r^2 {\mathcal F}(r)\quad ,
\eeqy
where $\mathcal{F}(r)$ is a functional of the nucleon density $n_q(r)$ and of 
the electron density $n_e(r)$. These densities are related to the total numbers
of nucleons and electrons in the cell by 
\bmlet
\beqy\label{B4a}
N_q=4\pi \int_0^R {\rm d}r\, r^2 n_q(r)\quad ,
\eeqy 
\beqy\label{B4b}
N_e=4\pi \int_0^R {\rm d}r\, r^2 n_e(r)\quad .
\eeqy 
\emlet
Combining Eqs.~(\ref{B2}) and (\ref{B3}) yields 
\beqy\label{B5}
P=-\mathcal{F}(R)- \frac{1}{R^2}\int_0^R{\rm d}r\, r^2 
\biggl(\sum_q \frac{\delta F}{\delta n_q(r)}\frac{\partial n_q(r)}{\partial R}
+\frac{\delta F}{\delta n_e(r)}\frac{\partial n_e(r)}{\partial R}\biggr)\quad ,
\eeqy
where $\delta F/\delta n_q(r)$ and $\delta F/\delta n_e(r)$ denote the 
functional derivatives of $F$ with respect to the nucleon and electron 
densities, respectively. 

Minimizing now the Helmholtz free energy $F$ with respect to arbitrary 
variations in $n_q(r)$ and $n_e(r)$ leads to the Euler-Lagrange equations
\bmlet
\beqy\label{B6a}
\lambda_q=\frac{\delta F}{\delta n_q(r)}
\eeqy
and
\beqy\label{B6b}
\lambda_e= \frac{\delta F}{\delta n_e(r)}\quad ,
\eeqy
\emlet
where the $\lambda_q$ and $\lambda_e$ are Lagrange multipliers introduced to 
ensure that the nucleon and electron numbers given by Eqs.~(\ref{B4a}) and 
(\ref{B4b}) remain fixed; they are identified with 
the corresponding chemical potentials. Using next the identities
\bmlet
\beqy\label{B7a}
\int_0^R {\rm d} r\, r^2 \frac{\partial n_q(r)}{\partial R}=-R^2 n_q(R)\quad ,
\eeqy
and
\beqy\label{B7b}
\int_0^R {\rm d} r\, r^2 \frac{\partial n_e(r)}{\partial R}=-R^2 n_e(R)\quad ,
\eeqy
\emlet
which follow from the differentiation of Eqs.~(\ref{B4a}) and (\ref{B4b}), 
respectively, we arrive at 
\beqy\label{B8}
P=-\mathcal{F}(R)+\lambda_e n_e(R)+\sum_q \lambda_q n_q(R)\, .
\eeqy
This pressure formula is a generalization of the expression derived in atomic 
physics in the framework of the Thomas-Fermi-Dirac model (see, e.g., 
Ref.~\cite{abs90} and references therein).

We decompose now the total Helmholtz free energy density in the WS cell into a 
nuclear part, a purely kinetic electron part and a Coulomb part,
\beqy\label{B9}
\mathcal{F}(r)=\mathcal{F}_{\rm nuc}(r)+\mathcal{F}_e(r)+
\mathcal{F}_{\rm Coul}(r)\quad .
\eeqy
Then substituting Eq.~(\ref{B9}) into Eq.~(\ref{B8}) leads to
\beqy\label{B10}
P=-\mathcal{F}_{\rm nuc}(R) - \mathcal{F}_e - \mathcal{F}_{\rm Coul}(R)
+ \lambda_e n_e + \sum_q \lambda_q n_q(R)\quad ,
\eeqy
where we are assuming that $n_e$ and $\mathcal{F}_e$ are position-independent.
We now examine in more detail the different components of $\mathcal{F}(r)$ 
appearing in Eq.(\ref{B9}).

In the fourth-order ETF method with Skyrme functionals the nuclear part
${\mathcal F}_{\rm nuc}(r)$ is a local functional of the nucleon densities 
$n_q(r)$ and their derivatives up to just the second order, provided the
higher-order terms have been integrated as described in Sec.~\ref{model}. 
Note that we have {\it not} included the proton-proton Coulomb interaction in 
${\mathcal F}_{\rm nuc}(r)$. As for
the electron gas, since it is supposed to be uniform we can write simply
\beqy\label{B11}
F_e=V {\mathcal F}_e(n_e,T)\quad ,
\eeqy
where ${\mathcal F}_e$ is the electron Helmholtz free-energy density, which
depends only on the electron density $n_e = N_p/V = \bar{n}_p$ and the
temperature $T$. The Coulomb part of the Helmholtz free energy is given by
\beqy\label{B12}
F_{\rm Coul} = F_{\rm Coul,dir} + F_{\rm Coul,ex} =  4\pi \int_0^R {\rm d} r\, r^2 
\biggl[\mathcal{F}_{\rm Coul,dir}(r) + \mathcal{F}_{Coul,ex}(r)\biggr] \quad .
\eeqy
Here the direct term is 
\beqy\label{B13}
\mathcal{F}_{\rm Coul,dir}(r) = \frac{e}{2} n_c(r) \phi(r) \quad ,
\eeqy
where $n_c(r)\equiv n_p(r)-n_e$ is the net electric-charge density, and 
$\phi(r)$
is the Coulomb potential, found on solving Poisson's equation to be given by 
\beqy\label{B14}
\phi(r)=4\pi e\int_0^{R} {\rm d} r^\prime\, r^{\prime 2} n_c(r^\prime)
\mathcal{K}(r,r^\prime)\quad ,
\eeqy
in which
\beqy\label{B15}
\mathcal{K}(r,r^\prime)=\frac{r+r^\prime-|r-r^\prime|}{2r r^\prime}\quad .
\eeqy
For $r = R$, Eq.~(\ref{B14}) reduces to
\beqy\label{B16}
\phi(R)=\frac{4\pi e}{R} \int_0^{R} {\rm d} r\, r^2 n_c(r) = 0\quad ,
\eeqy
the last step being a consequence of global charge neutrality. It then follows
from Eq.~(\ref{B13}) that
\beqy\label{B17}
\mathcal{F}_{\rm Coul,dir}(R) = 0  \quad .
\eeqy
For the Coulomb-exchange term we have
\beqy\label{B18}
\mathcal{F}_{\rm Coul,ex}(r) = -\frac{3e^2}{4}\biggl(\frac{3}{\pi}\biggr)^{1/3}
\biggl[x n_p(r)^{4/3}-\frac{1}{2}n_e^{4/3}\biggr] \quad ,
\eeqy
where  $x$ is usually equal to 1 but, as explained in Section~\ref{model}, is 
set equal to zero for the BSk forces of this paper; for the electrons we
have taken the extreme relativistic expression~\cite{sal61}. Then
\beqy\label{B19}
\mathcal{F}_{\rm Coul}(R) = -\frac{3e^2}{4}\biggl(\frac{3}{\pi}\biggr)^{1/3}
\biggl[x n_p(R)^{4/3}-\frac{1}{2}n_e^{4/3}\biggr] \quad .
\eeqy

To proceed we have to evaluate the chemical potentials appearing in 
Eq.~(\ref{B10}). The Euler-Lagrange equation~(\ref{B6a}) for nucleons can 
be written explicitly as
\beqy\label{B20}
\lambda_q = \frac{\partial {\mathcal F}_{\rm nuc}(r)}{\partial n_q(r)}-
\nabla\cdot\frac{\partial {\mathcal F}_{\rm nuc}(r)}{\partial \nabla n_q(r)}+
\nabla^2\frac{\partial {\mathcal F}_{\rm nuc}(r)}{\partial \nabla^2 n_q(r)}
+ \biggl[e\phi(r)-xe^2\left(\frac{3}{\pi}\right)^{1/3}n_p(r)^{1/3}\biggr]
\delta_{q,p} \, .
\eeqy
The constant $\lambda_q$ can be evaluated at any point $r \le R$, but taking
$r = R$ leads to a considerable simplification of the right-hand side of 
Eq.~(\ref{B20}), since with our parametrization all derivatives of the density 
vanish at that point. Thus the second and third terms of this expression 
likewise vanish at that point, since each can be expresssed as a sum of terms 
every one of which contains a factor of some derivative of $n_q(r)$. 
Using then Eq.~(\ref{B16}) the nucleon chemical potential becomes
\beqy\label{B21}
\lambda_q = \frac{\partial {\mathcal F}_{\rm nuc}(R)}{\partial n_q(R)} 
- xe^2\left(\frac{3}{\pi}\right)^{1/3}n_p(R)^{1/3} \delta_{q,p} \quad .
\eeqy 
A further consequence of the vanishing of the derivatives of $n_q(r)$ at $r=R$ 
is that the first term here, like the term ${\mathcal F}_{\rm nuc}(R)$ 
appearing in Eq. ~(\ref{B10}), involves only the bulk part of the nuclear 
free energy density.
Next, the Euler-Lagrange equation~(\ref{B6b}) for electrons simplifies to
\beqy\label{B22}
\lambda_e = \frac{\partial \mathcal{F}_e}{\partial n_e} - e\phi(r)+ 
\frac{e^2}{2} \left(\frac{3}{\pi}\right)^{1/3}n_e^{1/3} \quad ,
\eeqy
because of the uniformity of the electron gas. For the same reason we can write
the electron pressure (without the Coulomb exchange term) as
\beqy\label{B23}
P_e=-\frac{\partial F_e}{\partial V}=-{\mathcal F}_e+n_e
\frac{\partial {\mathcal F}_e}{\partial n_e} \quad .
\eeqy
Also, the Coulomb-potential term $e\phi(r)$ in Eq.~(\ref{B22}) vanishes at 
$r = R$, and must be negligible for $r < R$, since otherwise $n_e$ and 
${\mathcal F}_e$ would be position-dependent, which would be inconsistent with 
the assumption made and justified in Section~\ref{model} that the electron gas 
is essentially uniform in the inner crust. Then Eq.~(\ref{B22}) can be 
rewritten as
\beqy\label{B24}
\lambda_e\,n_e = P_e+\mathcal{F}_e +
\frac{e^2}{2} \left(\frac{3}{\pi}\right)^{1/3}n_e^{1/3} \quad .
\eeqy

Substituting now Eqs.~(\ref{B19}), (\ref{B21}) and (\ref{B24}) into
Eq.~(\ref{B10}) gives us for the total pressure 
\beqy\label{B25}
P = P_{\rm nuc}+P_e+P_{\rm Coul,ex} \quad ,
\eeqy
where 
\beqy\label{B26}
P_{\rm nuc}= -{\mathcal F}_{\rm nuc}(R)+ \sum_q
n_q(R)\frac{\partial {\mathcal F}_{\rm nuc}(R)}{\partial n_q(R)} \quad 
\eeqy
and
\beqy\label{B27}
P_{\rm Coul,ex}=\frac{e^2}{8}\left(\frac{3}{\pi}\right)^{1/3}n_e^{4/3}
-x\frac{e^2}{4}\left(\frac{3}{\pi}\right)^{1/3}n_p(R)^{4/3} \quad .
\eeqy
Given that both terms on the right-hand side of Eq.~(\ref{B26}) relate only to bulk 
matter, being independent of any density-gradient terms, it is easy to show from 
Eq.~(\ref{B1}) that $P_{\rm nuc}$ represents the purely nuclear pressure of 
homogeneous nuclear matter 
with neutron and proton densities equal to $n_n(R)$ and $ n_p(R)$, respectively, 
without any Coulomb term, direct or exchange. However, from Eq.~(\ref{B18}) it is 
seen that the last term of Eq.~(\ref{B27}) is just the Coulomb exchange pressure 
associated with the protons of this homogeneous system, while the first term of 
this equation
is likewise the Coulomb exchange pressure of the uniform electron gas. 

This means that the pressure of any crustal layer is the same as that obtained in a homogeneous
medium of neutrons, protons and electrons, with the neutron and proton densities being
those found at the surface of the WS cell, i.e., in the homogeneous background,
$n_{Bn}$ and $n_{Bp}$, respectively, while the electron density is to be taken as 
that of the actual uniform electron gas, $n_e$. It is remarkable that the direct 
Coulomb contribution, calculated exactly, vanishes identically,
even though $n_p(R)$ is not equal to $n_e$. However, this term still manifests itself
indirectly, since it influences the actual values of $n_n(R)$ and
$n_p(R)$ through the Euler-Lagrange equations. A similar remark applies also
to the inhomogeneities inside the cell.

For the generalized Skyrme force~(\ref{1.1}), the purely nuclear pressure can be 
expressed as
\beqy\label{B29}
P_{\rm nuc} &=& \frac{\hbar^2}{3 M}\tau_0+\sum_{t=0,1}\biggl( C_t^n n_{Bt}^2+\frac{5}{3}C_t^\tau n_{Bt}\tau_t
+n_{B0}\frac{\partial C_t^n}{\partial n_{B0}} n_{Bt}^2+n_{B0}\frac{\partial C_t^\tau}{\partial n_{B0}} n_{Bt}\tau_t\biggr)\quad ,
\eeqy
where $n_{B0}=n_{Bn}+n_{Bp}$, while $n_{B1}=n_{Bn}-n_{Bp}$, and likewise for 
$\tau_0$ and $\tau_1$, with 
\beqy\label{B30}
\tau_q=\frac{3}{5}(3\pi^2)^{2/3}n_q(R)^{5/3}\quad .
\eeqy
The various coefficients are given by 
\bmlet
\beqy
C_0^n=\frac{3}{8}t_0+\frac{3}{48}t_3 n_{B0}^\alpha
\eeqy
\beqy
C_1^n=-\frac{1}{4}t_0\left(\frac{1}{2}+x_0\right)-\frac{1}{24}t_3(1+x_3)n_{B0}^\alpha
\eeqy
\beqy
C_0^\tau=\frac{3}{16}t_1+\frac{1}{4}t_2\left(\frac{5}{4}+x_2\right)+\frac{3}{16}t_4 n_{B0}^\beta +\frac{1}{4}t_5\left(\frac{5}{4}+x_5\right)n_{B0}^\gamma
\eeqy
\beqy
C_1^\tau=-\frac{1}{8}t_1\left(\frac{1}{2}+x_1\right)+\frac{1}{8}t_2\left(\frac{1}{2}+x_2\right)-\frac{1}{8}t_4 n_{B0}^\beta \left(\frac{1}{2}+x_4\right)+\frac{1}{8}t_5n_{B0}^\gamma\left(\frac{1}{2}+x_5\right)\quad .
\eeqy
\emlet
The pressure $P_e$ of the uniform electron gas is calculated as described in
Section~\ref{model}, using expressions given in Section 24 of Cox and
Giuli~\cite{cg04}.

\begin{acknowledgments}
We wish to thank M. Brack for helpful comments.
This work was financially supported by the NSERC (Canada), the FNRS (Belgium),
the Communaut\'e fran\c{c}aise de Belgique (Actions de Recherche Concert\'ees),
and CompStar (a Research Networking Programme of the European Science
Foundation).
\end{acknowledgments}


\begin{thebibliography}{99}
\bibitem{pr95}C. J. Pethick and D. G. Ravenhall, Ann. Rev. Nucl. Part. Sci.
{\bf 45}, 429 (1995).
\bibitem{lrr} N. Chamel and P. Haensel,``Physics of Neutron Star Crusts'', 
Living Rev. Relativity {\bf 11}, (2008), 10. 
http://www.livingreviews.org/lrr-2008-10
\bibitem{gcp10}S. Goriely, N. Chamel, and J.~M. Pearson,
Phys. Rev. C {\bf 82}, 035804 (2010).
\bibitem{cha10} N. Chamel, Phys. Rev. C {\bf 82}, 014313 (2010).
\bibitem{cg10} N. Chamel and S. Goriely, Phys. Rev. C {\bf 82}, 045804 (2010).
\bibitem{audi03} G. Audi, A.H. Wapstra, and C. Thibault,
Nucl. Phys. \textbf{A729}, 337  (2003).
\bibitem{pgc11}J.~M. Pearson, S. Goriely, and N. Chamel, Phys. Rev. C {\bf 83}, 065810 (2011).
\bibitem{cha98}E. Chabanat, P. Bonche, P. Haensel, J. Meyer, and R. Schaeffer,
Nucl. Phys. {\bf A635}, 231 (1998); Nucl. Phys. {\bf A643}, 441 (1998).
\bibitem{sg05}S. Goriely, M. Samyn, J.~M. Pearson, and M. Onsi,
Nucl. Phys. {\bf A750}, 425 (2005).
\bibitem{dsn04}J. Dobaczewski, M.V. Stoitsov, and W. Nazarewicz,
AIP Conference Proceedings Volume 726, ed. R. Bijker, R.F. Casten, A. Frank
(American Institute of Physics, New York, 2004) p. 51.
\bibitem{nv73}J. W. Negele and D. Vautherin, Nucl. Phys. {\bf A207}, 298 (1973).
\bibitem{marg07}J. Margueron, N. van Giai, and N. Sandulescu,
Proceedings of the International Symposium EXOCT07 (edited by U. Lombardo,
M. Baldo, F. Burgio, and H.-J. Schulze), p. 362 (2007).
\bibitem{cha07}N. Chamel, S. Naimi, E. Khan, and J. Margueron,
Phys. Rev. C {\bf 75}, 055806 (2007).
\bibitem{gri11}F. Grill, J. Margueron, and N. Sandulescu,
Phys. Rev. C {\bf 84}, 065801 (2011).
\bibitem{mag02} P. Magierski and P.-H. Heenen, Phys. Rev. C {\bf 65}, 045804 (2002).
\bibitem{gog07} P. G\"ogelein and H. M\"uther, 
Phys. Rev. C {\bf 76}, 024312 (2007).
\bibitem{new09} W.G. Newton and J.R. Stone, Phys. Rev. C {\bf 79}, 055801 (2009).
\bibitem{dh01}F. Douchin and P. Haensel, Astron. and Astrophys. {\bf 380},
151 (2001). 
\bibitem{oya07} K. Oyamatsu and K. Iida, Phys. Rev. C{\bf 75}, 015801 (2007). 
\bibitem{ons08}M. Onsi, A.~K. Dutta, H. Chatri, S. Goriely, N. Chamel, and
J.~M. Pearson, Phys. Rev. C {\bf 77} 065805 (2008).
\bibitem{ons97} M. Onsi, H. Przysiezniak, and J. M. Pearson, Phys. Rev. C{\bf 55}, 
3139 (1997).
\bibitem{oy94}K. Oyamatsu and M. Yamada, Nucl. Phys. {\bf A578}, 181 (1994).
\bibitem{ch06}N. Chamel, Nucl. Phys. {\bf A773}, 263 (2006).
\bibitem{tol39}R. C. Tolman, Phys. Rev. {\bf 55}, 364 (1939).
\bibitem{ov39}J. R. Oppenheimer and G. M. Volkoff, Phys. Rev. {\bf 55}, 374
(1939).
\bibitem{cgp09}N. Chamel, S. Goriely, and J. M. Pearson,
Phys. Rev. C {\bf 80}, 065804 (2009).
\bibitem{bgh85}M. Brack, C. Guet, and H. -B. H\aa kansson,
Phys. Reports {\bf 123}, 275 (1985).
\bibitem{wat03} G. Watanabe and K. Iida, Phys. Rev. C {\bf 68}, 045801 (2003).
\bibitem{mar05} T. Maruyama, T. Tatsumi, D. N. Voskresensky, T. Tanigawa, and 
S. Chiba, Phys. Rev. C {\bf 72}, 015802 (2005).
\bibitem{cg04}A. Weiss, W. Hillebrandt, H.-C. Thomas, and H. Ritter,
{\it Cox and Giuli's Principles of Stellar Structure}, extended second edition.
Cambridge Scientific Publishers (2004).
\bibitem{gp08}S. Goriely and J. M. Pearson, 
Phys. Rev. C  \textbf{77}, 031301(R) (2008).
\bibitem{sal61}E. Salpeter, ApJ. {\bf 134}, 669 (1961).
\bibitem{duc07}C. Ducoin, Ph. Chomaz, and F. Gulminelli,
Nucl. Phys. {\bf A789}, 403 (2007).
\bibitem{dh00}F. Douchin and P. Haensel, Phys. Lett. B {\bf 485}, 107 (2000).
\bibitem{hempel}http://phys-merger.physik.unibas.ch/~hempel/eos/oc/sly4.eos
\bibitem{rhs06}S. B. R\"uster, M. Hempel, and J Schaffner-Bielich,
Phys. Rev. C {\bf 73}, 035804 (2006).
\bibitem{st83}S. L. Shapiro and S. A. Teukolsky, {\it Black Holes, White
Dwarfs, and Neutron Stars}, Wiley Interscience (1983).
\bibitem{abs90} A.M. Abrahams and S. L. Shapiro, Phys. Rev. A {\bf 42}, 2530 
(1990).  
\bibitem{cha11}N. Chamel, A. F.Fantina, J. M. Pearson, and S. Goriely,
Phys. Rev. C {\bf 84}, 062802 (2011). 
\end{thebibliography}
\end{document}